\documentclass[acmsmall,screen]{acmart}

\newcommand{\xqym}[1]{#1}

\setcopyright{acmlicensed}
\copyrightyear{2018}
\acmYear{2018}
\acmDOI{XXXXXXX.XXXXXXX}
\acmConference[Conference acronym 'XX]{Make sure to enter the correct
  conference title from your rights confirmation email}{June 03--05,
  2018}{Woodstock, NY}
\acmISBN{978-1-4503-XXXX-X/2018/06}

\usepackage[dvipsnames]{xcolor}
\usepackage{hyperref}

\definecolor[named]{ACMPurple}{cmyk}{0.55,1,0,0.15}
\hypersetup{
    colorlinks=true,
    linkcolor=ACMPurple,
    filecolor=ACMPurple,
    citecolor=ACMPurple,
    urlcolor=cyan,
    pdftitle={A Minimalist Proof Langauge},
    pdfpagemode=FullScreen,
    }

\usepackage{subcaption}

\usepackage{amsmath}
\usepackage{amsthm}
\usepackage{algorithmic}
\usepackage{array}

\usepackage{url}

\usepackage{enumitem}

\newlist{questions}{enumerate}{2}
\setlist[questions,1]{label=\textbf{RQ \arabic*:},ref=RQ \arabic*}
\setlist[questions,2]{label=(\alph*),ref=\thequestionsi(\alph*)}

\usepackage{cleveref}
\usepackage{listings}
\usepackage{tikz}
\usetikzlibrary{positioning,decorations.markings,arrows.meta}
\usepackage[most]{tcolorbox}
\usepackage[T1]{fontenc}
\usepackage{mathtools}
\usepackage{booktabs}
\usepackage[edges]{forest}
\usepackage{wrapfig}

\newsavebox{\codebox}
\newsavebox{\reorgboxA}
\newsavebox{\reorgboxB}
\newsavebox{\reorgboxC}
\newsavebox{\reorgboxD}
\newsavebox{\reorgboxE}
\newsavebox{\subgoalBOX}
\newsavebox{\subgoalBOXresidue}
\newsavebox{\bracketBOX}
\newsavebox{\bracketBOXB}
\newsavebox{\transBoxA}
\newsavebox{\transBoxB}
\newsavebox{\transBoxC}
\newsavebox{\transBoxD}
\newsavebox{\transBoxE}
\newsavebox{\transBoxF}
\newsavebox{\transBoxG}
\newsavebox{\transBoxH}

\newcommand{\translate}[1]{\left\llbracket#1\right\rrbracket}

\crefname{part}{\S\!}{\S\S}
\crefname{chapter}{\S\!}{\S\S}
\crefname{section}{\S\!}{\S\S}
\crefname{subsection}{\S\!}{\S\S}
\crefname{figure}{Fig.}{Figs.}
\crefname{table}{Table}{Tables}
\Crefname{table}{Table}{Tables}

\definecolor{dkgreen}{rgb}{0,.6,0}
\definecolor{dkblue}{rgb}{0,0,.6}
\definecolor{dkyellow}{cmyk}{0,0,.8,.3}

\definecolor{keywordcolor}{rgb}{0.7, 0.1, 0.1}   % red
\definecolor{tacticcolor}{rgb}{0.0, 0.1, 0.6}    % blue
\definecolor{commentcolor}{rgb}{0.4, 0.4, 0.4}   % grey
\definecolor{symbolcolor}{rgb}{0.0, 0.1, 0.6}    % blue
\definecolor{sortcolor}{rgb}{0.1, 0.5, 0.1}      % green
\definecolor{attributecolor}{rgb}{0.7, 0.1, 0.1} % red

\definecolor{isarblue}{HTML}{006699}
\definecolor{MLblue}{HTML}{00334D}
\definecolor{isarlight}{HTML}{0099FF}
\definecolor{isargreen}{HTML}{009966}
\definecolor{isarpurple}{HTML}{800080}
\lstdefinelanguage{isar}{%
    keywords=[1]{type_synonym,datatype,fun,abbreviation,definition,lemma,theorem,corollary,have,by,obtain,consider,let,\{,\},apply,apply_end,done,subgoal,define,write,interpret,note},
    keywordstyle=[1]\bfseries\color{isarblue},
    keywords=[2]{where,assumes,shows,and,if,for,premises},
    keywordstyle=[2]\bfseries\color{isargreen},
    keywords=[3]{else,case,of,SOME,in,O,proof,qed,show,assume,fix,next},
    keywordstyle=[3]\bfseries\color{isarlight},
    keywords=[4]{with,then,thus,hence,from,using,also,finally,moreover,ultimately},
    keywordstyle=[4]\bfseries\color{isarblue},
showstringspaces=false,
keepspaces=true,
columns=[l]flexible,
}

\lstdefinelanguage{MiniLang}{%
    keywords=[1]{theorem,lemma},
    keywordstyle=[1]\bfseries\color{isarblue},
    keywords=[2]{WITH,WITHOUT,where,and,VARS,arbitrary,rule},
    keywordstyle=[2]\bfseries\color{isargreen},
    keywords=[3]{RULE,LET,CONSIDER,UNFOLD,CASE_SPLIT,INDUCT,CHOOSE,APPLY,INTRO,HAVE,END,NEXT,SIMP,CONFIG,OPEN,SIMPLIFY,NOTATION},
    keywordstyle=[3]\bfseries\color{MLblue},
columns=[l]flexible,
}

\newcommand{\isar}[1]{\lstinline[language=isar]{#1}}
\newcommand{\minilang}[1]{\lstinline[language=MiniLang]{#1}}

\usepackage{multirow}

\newcommand{\CCTX}[1]{#1}
\newcommand{\CGOAL}[1]{#1}
\newcommand{\NOTE}[1]{#1}

\makeatletter
\def\namedlabel#1#2{\begingroup
    #2%
    \def\@currentlabel{#2}%
    \phantomsection\label{#1}\endgroup
}
\makeatother

\makeatletter
\def\namedLabel#1#2#3{\begingroup
    #3%
    \def\@currentlabel{#2}%
    \phantomsection\label{#1}\endgroup
}
\makeatother

\makeatletter
  \newskip\OrigTopsep
  \newskip\OrigPartopsep
  \newskip\OrigParskip
\makeatother

\OrigTopsep    = \topsep
\OrigPartopsep = \partopsep
\OrigParskip   = \parskip

\newtheorem{theorem}{Theorem}

\newenvironment{finding}{%
  \begin{tcolorbox}[
    enhanced,breakable,
    colback=gray!20,colframe=gray!20,
    arc=3mm,boxrule=0pt,
    left=6pt,right=6pt,top=4pt,bottom=4pt,
    before skip=5pt,
    after skip=5pt,
    fontupper=\itshape
  ]%
  \normalfont\color{black}
}{%
  \end{tcolorbox}%
}

\newcommand{\DS}[1]{}

\newcommand{\CW}[1]{}

\usetikzlibrary{calc}

\usepackage{comment}

\begin{document}

%%%
%%% The "title" command has an optional parameter,
%%% allowing the author to define a "short title" to be used in page headers.
\title{A Minimalist Proof Language for Neural Theorem Proving over Isabelle/HOL}
%\title{IsaMini: Redesigned Isabelle Proof Language for Machine Learning}
%
% \author{\IEEEauthorblockN{Qiyuan Xu\IEEEauthorrefmark{1},
% Renxi Wang\IEEEauthorrefmark{2},
% Haonan Li\IEEEauthorrefmark{2},
% David San\'an\IEEEauthorrefmark{3}, and
% Conrad Watt\IEEEauthorrefmark{1}}
% \IEEEauthorblockA{\IEEEauthorrefmark{1}%
% %College of Computing and Data Science,
% Nanyang Technological University,
% Singapore 639798.
% Email: \{first name\}.\{last name\}@ntu.edu.sg}
% \IEEEauthorblockA{\IEEEauthorrefmark{2}Mohamed bin Zayed University of Artificial Intelligence, Abu Dhabi, UAE.\\
% Email: \{first name\}.\{last name\}@.mbzuai.ac.ae}
% \IEEEauthorblockA{\IEEEauthorrefmark{3}Singapore Institute of Technology, Singapore 828608, Email: david.miguel@singaporetech.edu.sg}
% }

%%
%% The "author" command and its associated commands are used to define
%% the authors and their affiliations.
%% Of note is the shared affiliation of the first two authors, and the
%% "authornote" and "authornotemark" commands
%% used to denote shared contribution to the research.
\author{Qiyuan Xu}
\email{xu@qiyuan.me}
\orcid{0000-0002-9196-3237}
\affiliation{%
  \institution{Nangyang Technological University}
  \country{Singapore}
}

\author{Renxi Wang}
\orcid{}
\email{renxi.wang@mbzuai.ac.ae}
\affiliation{%
  \institution{Mohamed bin Zayed University of Artificial Intelligence}
  \city{Abu Dhabi}
  \country{United Arab Emirates}
}

\author{Peixin Wang}
\orcid{}
\email{pxwang@sei.ecnu.edu.cn}
\affiliation{%
  \institution{East China Normal University}
  \city{Shanghai}
  \country{China}
}

\author{Haonan Li}
\orcid{0000-0001-6623-5089}
\email{haonan.li@mbzuai.ac.ae}
\affiliation{%
  \institution{Mohamed bin Zayed University of Artificial Intelligence}
  \city{Abu Dhabi}
  \country{United Arab Emirates}
}

\author{Conrad Watt}
\orcid{0000-0002-0596-877X}
\email{conrad.watt@ntu.edu.sg}
\affiliation{%
  \institution{Nangyang Technological University}
  \country{Singapore}
}

%%
%% By default, the full list of authors will be used in the page
%% headers. Often, this list is too long, and will overlap
%% other information printed in the page headers. This command allows
%% the author to define a more concise list
%% of authors' names for this purpose.
\renewcommand{\shortauthors}{Qiyuan et al.}

\begin{abstract}
Neural Theorem Proving (NTP) employs Large Language Models (LLMs) to automate formal proofs in proof assistants.
While LLMs have achieved relatively remarkable success in informal reasoning tasks using natural languages, the transition to mechanized formal theorem proving presents persistent challenges.
Mechanized proof languages often contain many syntactic constructs and diverse, specialized proof tactics, which facilitate expert use but have no direct counterpart in informal mathematical proofs.
These prover-specific idioms represent an additional burden for LLM-based NTPs that might be otherwise successful in generating informal proofs.
% right now you don't say what the "gap" is - see talking point just below
%
%%%
%While these features enhance expert efficiency in proof
%construction, they introduce proof-assistant-specific idioms that have %no direct counterpart in
%informal mathematical proofs, thereby widening the gap between formal %and informal theorem
%proving and imposing learning burdens on language models.
%%%
%We identify common operations in informal reasoning and design a proof language that implements these operations.
%%% CONRAD: put this narrative
%
Seeking to bridge this gap between formal proof construction and informal reasoning, in order to better facilitate NTP, this work approaches these challenges from a language design perspective.
We look at common reasoning patterns in informal proofs and in existing mechanized proofs, and design \emph{Minilang} \xqym{(formally named Isabelle/Minilang),} a minimalist proof language that captures these reasoning patterns.
%\pw{What operations?}
% smallest language + alignment
%The result is \emph{Minilang}, a redesigned proof language for Isabelle/HOL incorporating an improved version of Sledgehammer.
In contrast to proof languages (informal and formal) that often feature a large collection of operations with unclear semantic boundaries, Minilang is deliberately kept minimalist --- its core design comprises only 10 proof operations, each with clear semantic distinctions.
We further develop a rule-based translator from Isabelle's proof language (Isar) to Minilang, translating ${\sim}340$K existing Isabelle proofs with an ${\sim}85\%$ success rate.
Using this translated corpus, we finetune two LLMs to compare machine learning performance on Minilang versus the original Isar language.
Experiments show Minilang benefits the two LLMs by improving the pass@1 success rate on the PISA benchmark by up to 20/29 percentage points in comparison to the Isar-based LLMs w/wo Sledgehammer.
The pass@1 rate reaches 69.1\%, exceeding the prior work Baldur's pass@64 (65.7\%); the pass@8 rate reaches 79.2\%, exceeding the state-of-the-art on PISA (71.0\%) achieved by Magnushammer.
\end{abstract}

\maketitle

\section{Introduction}\label{sec:introduction}

Formal verification of software systems fundamentally relies on theorem proving to certify safety-critical properties.
%with celebrated end-to-end results such as the seL4 microkernel verification~\cite{seL4}.
While interactive theorem proving (ITP) requires substantial manual effort and automated theorem proving (ATP) excels only within restricted domains, the rise of Large Language Models (LLMs) has introduced a promising alternative: Neural Theorem Proving (NTP). By using LLMs to interface with interactive theorem provers, NTP enables automated proof construction for complex properties~\cite{li2024survey}.
Recent NTP systems have demonstrated remarkable success in mathematical competitions~\cite{AlphaProver,SeedProver,Kimina}, achieving gold medal performance at the International Mathematical Olympiad~\cite{Aristotle}, indicating promising reasoning and theorem-proving capabilities.

%However, NTP performance in wider formal reasoning settings has been mixed. CITATION NEEDED ??
%\pw{Add some examples/literatures to support the pros and cons of NTP.}

%Nowadays, most NTP systems are based on declarative theorem proving, where language models construct formal proofs by leveraging their reasoning capabilities developed on natural language corpora.

%To narrow this gap, prior works primarily focus on model training techniques.
%In this work, we take a complementary approach by exploring the potential of proof language design.

%Our key hypothesis is that when proof languages mirror the operations naturally employed in informal proofs, the alignment between formal and informal reasoning becomes more transparent, thereby easing the learning process for language models and finally improving the performance of language model-based neural theorem proving.

A fundamental challenge in NTP is the gap between natural-language-based informal reasoning and formal-language-based theorem proving.
Prior works focus on training language models to master the existing languages of proof assistants~\cite{PALM, Baldur, DPSKv2, Kimina}.
%Existing NTP works focus on generating proof script in the unmodified language of an interactive theorem prover.
%
In this work, we take a complementary approach by exploring the potential of proof language design focused specifically on facilitating NTP.
%
%Our key hypothesis is that by minimizing the presence of prover-specific idioms and reducing the complexity of the proof language, the performance of language model-based neural theorem proving will improve.
Our key hypothesis is that the performance of language model-based neural theorem proving can be improved by reducing the prover-specific idioms and the complexity of the proof language.

%Admittedly, existing proof languages~\cite{Isar,Mizar,Lean} already incorporate declarative constructs that reflect natural proof operations, such as Isabelle's \texttt{consider} and Lean's \texttt{have}.
%However, these constructs are primarily motivated by enabling human mathematicians to write and read proofs comfortably, prioritizing expert productivity over model learnability.
%
%In particular, these languages offer a rich array of expert features for fine-grained control over the reasoning system, such as specialized reasoning mechanisms and extensive libraries of tactics that offer complex configurable parameters.

%While these features enhance expert efficiency in proof construction, they introduce proof-assistant-specific idioms that have no direct counterpart in informal mathematical proofs, thereby widening the gap between formal and informal theorem proving and imposing learning burdens on language models.

% CONRAD: the above should show up somewhere (Motivation)

We therefore propose \xqym{\emph{Isabelle/Minilang}}, a minimalist proof language that \emph{restricts its constructs to only those operations with natural counterparts in informal mathematical reasoning}.
An example is illustrated in \cref{fig:language}.
For the proof obligations raised from those operations, Minilang delegates them to ATPs and instead focuses on high-level proof structures, as language models excel at proof outlining rather than fine-grained deduction.
Its semantics is formalized using a simple tree-based state machine, which not only provides the foundation for establishing its soundness but also demonstrates the language's conceptual simplicity at the semantic level, maintaining consistency with its minimalist design philosophy.
%\pw{Why MinLang is minimal? Any limitations of it, e.g., the expressivity is lower? If not, claim this.}

We also present \emph{Sledgehammer*}, the ATP used by Minilang, which is an improved version of Sledgehammer. Our improvements lie in two aspects.
First, Sledgehammer* incorporates a preprocessing step that simplifies and decomposes proof goals into subgoals more amenable to automated proving.
Second, we train our proof generation models to produce relevant lemma hints to guide Sledgehammer*'s premise selection, allowing the models' global proof strategy to directly complement the hammer's local focus on individual subgoals.

Beyond this language design, we further contribute a rule-based translator to convert existing Isabelle corpora into Minilang for model training.
This is challenging due to Isabelle's complex syntax and numerous corner cases. We propose three strategies to overcome this complexity:
%\begin{itemize}
%\item
(i) \emph{Elaboration}: Make implicit information explicit by exposing hidden details in a clear, structured manner;
%\item
(ii) \emph{Normalization}: Consolidate diverse approaches for achieving the same logical purpose into uniform representations;
%\item
(iii) \emph{Elimination of tactics}: Replace tactics with Sledgehammer*, except those that correspond well to informal proofs.
%\end{itemize}
Applying this translator at scale, 
%Applying it at scale,
we translate 85.25\% of ${\sim}340$K existing Isar proofs into Minilang.
This translation distills Isar's versatile syntax and idiomatic reasoning into unified minimalist proof structures, eliminating unnecessary variations and theorem-proving idioms for more efficient LLM learning.

To quantify the gains from this representation distillation, we fine-tune two pretrained LLMs --- Llemma~\cite{Llemma} and Deepseek Prover Base v1.5~\cite{DPSK1.5} --- on both the translated Minilang corpus and the original Isar corpus for comparison.
The results show that Minilang models achieve about a 29 percentage point improvement over Isar models.

Since Minilang incorporates ATP to discharge proof obligations while Isar does not, we conduct an ablation study to isolate the impact of language redesign from the ATP enhancement.
As our baseline, we adopt the currently strongest known method, Thor's approach~\cite{Thor}. It augments Isar scripts by replacing tactics with Sledgehammer calls wherever applicable.
Even against this stronger baseline, Minilang maintains a 20 percentage point advantage, demonstrating the effectiveness of its minimalist language design.

%Premises selection is Sledgehammer's core function, which identifies from the system database the relevant lemmas that may help to prove a given goal.
% Move this to related works
%Although the adoption of hammers in NTP dates back to Thor~\cite{Thor}, to our knowledge, all prior works ignore that Sledgehammer can accept hints of relevant lemmas to improve its premise selection.
%We argue that training NTP models to generate these hints can produce a significant improvement because it enables the proof generation models to suggest relevant lemmas based on the models' global proof strategy, complementing hammers' local focus on individual subgoals.
To measure the improvement provided by Sledgehammer* and our integration of relevant lemma hints, we conduct an ablation study that compares Sledgehammer* against the original Sledgehammer on both Minilang and Isar. The result shows at least 6 percentage points of improvement across all our metrics.
%The second experiment compares our approach against the prior state-of-the-art, Magnushammer. The result shows ...

\noindent\textbf{Contributions.} To conclude this introduction, we summarize our contributions:
\begin{itemize}
\item We present Minilang, a minimalist proof language over Isabelle, designed to restrict language constructs to operations that align with informal proofs, to better support language model-based neural theorem proving.
%Experiments show ${\sim}20$ percentage point improvement over Isar-based LLMs with Sledgehammer.

\item We develop a rule-based translator from Isar to Minilang that successfully converts 85.28\% of 340K existing Isar proofs.
\item We present two fine-tuned whole-proof generation models over Minilang, respectively based on Llemma and Deepseek-Prover Base v1.5.
\item We present Sledgehammer*, an improved Sledgehammer that additionally takes relevant remise hints as input. We further provide two fine-tuned language models for generating these hints, constituting two automatic hammers. %Experiment shows ...
\item We develop an open-source socket-based Isabelle Read-Eval-Print-Loop infrastructure that scales to large compute clusters. We have deployed it on a cluster of $\text{24 nodes} \times \text{128 CPUs}$.
\end{itemize}

\noindent\textbf{Outline.} The remainder of this paper is organized as follows. We begin by presenting our motivation in~\cref{sec:motivation}. In~\cref{sec:minilang}, we introduce Minilang, a minimalist proof language, along with Sledgehammer*, the enhanced ATP system it employs. We then propose an automated translation from Isar to Minilang in~\cref{sec:translation}, which provides a substantial proof corpus for NTP training. Our experimental results are presented in~\cref{sec:evaluation}, followed by a discussion of related work in~\cref{sec:related-works}. Finally, we conclude with the limitations and future work in~\cref{sec:conclusion}.

\makeatletter
\newenvironment{btHighlight}[1][]
{\begingroup\tikzset{bt@Highlight@par/.style={#1}}\begin{lrbox}{\@tempboxa}}
{\end{lrbox}\bt@HL@box[bt@Highlight@par]{\@tempboxa}\endgroup}

\newcommand\btHL[1][]{%
  \begin{btHighlight}[#1]\bgroup\aftergroup\bt@HL@endenv%
}
\def\bt@HL@endenv{%
  \end{btHighlight}%   
  \egroup
}
\newcommand{\bt@HL@box}[2][]{%
  \tikz[#1]{%
    \pgfpathrectangle{\pgfpoint{1pt}{0pt}}{\pgfpoint{\wd #2}{\ht #2}}%
    \pgfusepath{use as bounding box}%
    \node[anchor=base west, fill=orange!30,outer sep=0pt,inner xsep=1pt, inner ysep=0pt, rounded corners=3pt, minimum height=\ht\strutbox+1pt,#1]{\raisebox{1pt}{\strut}\strut\usebox{#2}};
  }%
}
\makeatother

\lstdefinestyle{XXX}{
basicstyle=\ttfamily\footnotesize,
%moredelim=**[is][{\btHL[fill=yellow!15]}]{@}{@},
moredelim=**[is][]{@}{@},
%moredelim=**[is][\btHL]{!}{!},
escapeinside={&}{&}
}

\lstdefinestyle{YYY}{
basicstyle=\ttfamily\small,
moredelim=**[is][{\btHL[fill=yellow!15]}]{@}{@},
%moredelim=**[is][\btHL]{!}{!},
escapeinside={&}{&}
}

%\newcommand{\btHL}[1]{%
%  \sbox0{#1}%
%  \begin{tikzpicture}[baseline=(box.base)]
%    \node[fill=yellow!30,inner sep=1pt,rounded corners=2pt] (box) {\usebox0};
%  \end{tikzpicture}%
%}
%\setlength{\fboxsep}{5pt}
%\lstset{columns=fullflexible,keepspaces=true}
%\newcommand{\codehl}[1]{\colorbox{yellow}{#1}}
\begin{figure}[p]

\centering
\begin{subfigure}[b]{0.48\linewidth}
\begin{lstlisting}[language=coq, style=XXX]
Theorem sqrt2_not_rational:
  forall pq : nat, q <> 0 -> p*p = 2*(q*q) -> False.
@intros p q; generalize p; clear p;@
@elim q using (well_founded_ind lt_wf).@
@clear q; intros q Hrec p Hneq; @
@pose proof Hneq as Hlt_O_q;@
@apply Nat.neq_0_lt_0 in Hlt_O_q;@
@intros Heq.@
@apply (Hrec (3 * q - 2 * p)@
            @(comparison4 _ _ Hlt_O_q Heq)@
            @(3 * p - 4 * q)).@
@apply sym_not_equal; apply lt_neq;@
@apply Nat.add_lt_mono_l with (2 * p);@
@rewrite <- plus_n_O; rewrite Nat.add_comm;@
@rewrite Nat.sub_add; auto with *.@
@apply new_equality; auto.@
Qed.
\end{lstlisting}
\caption{Rocq, from~\cite{coq-sqrt2}}
\label{fig:pure-tactic-proof}
\end{subfigure}\hspace*{0.04\linewidth}%
\begin{subfigure}[b]{0.48\linewidth}
\begin{lstlisting}[language=lean, style=XXX]
theorem sqrt_two_irrational {a b : ℕ}
  (co : gcd a b = 1) : a^2 ≠ 2 * b^2 :=
by @rintro h : a^2 = 2 * b^2@
  have : 2 | a^2 := by @simp [h]@
  have : 2 | a := @dvd_of_dvd_pow prime_two this@
  @apply Exists.elim this@ @rintro c aeq@
  have : 2 * (2 * c^2) = 2 * b^2 := by
    @simp [Eq.symm h, aeq]@
    @simp [pow_succ' _, mul_comm, ...]@
  have : 2 * c^2 = b^2 := by
    @apply mul_left_cancel₀ _ this@ @decide@
  have : 2 | b^2 := by @simp [Eq.symm this]@
  have : 2 | b := by
    @exact dvd_of_dvd_pow prime_two this@
  have : 2 | gcd a b := by
    @apply dvd_gcd . assumption . assumption@
  have _ : 2 | (1 : ℕ) := by @simp [co] at *@
  @contradiction@
\end{lstlisting}
\caption{Lean, from~\cite{lean-sqrt2}.}
\label{fig:lean-sqrt2}
\end{subfigure}

\vspace{1em}
\begin{subfigure}[b]{0.5\linewidth}
\begin{lstlisting}[language=isar,
style=XXX,
%moredelim={**[is][{\btHL[fill=cyan!15]}]{@}{@}}
]
theorem sqrt2_not_rational: "sqrt 2 &$\notin\;\mathbb{Q}$&"
proof
  let ?x = "sqrt 2"
  assume "?x &$\in\;\mathbb{Q}$&"
  then obtain m n :: nat where
    "|?x| = m / n" and "coprime m n"
    @by (rule Rats_abs_nat_div_natE)@
  hence "m^2 = ?x^2 * n^2"
    @by (auto simp add: power2_eq_square)@
  hence eq: "m^2 = 2 * n^2"
    @using of_nat_eq_iff power2_eq_square by force@
  hence "2 dvd m^2" @by simp@
  hence "2 dvd m" @by simp@
  have "2 dvd n" proof -
    from "2 dvd m" obtain k where "m = 2 * k" @..@
    with eq have "2 * n^2 = 2^2 * k^2" @by simp@
    hence "2 dvd n^2" @by simp@
    thus "2 dvd n" @by simp@
  qed
  with "2 dvd m" have "2 dvd gcd m n"
    @by (rule gcd_greatest)@
  with lowest_terms have "2 dvd 1" @by simp@
  thus False @using odd_one by blast@
qed
\end{lstlisting}
\vspace{-0.7em}
\caption{Isabelle/Isar, from~\cite{isar-sqrt2}}
\label{fig:isar-sqrt2}
\end{subfigure}\hspace{0.035\linewidth}%
\begin{subfigure}[b]{0.465\linewidth}
\begin{lstlisting}[language=MiniLang,
style=XXX,
moredelim={**[is][{\btHL[fill=cyan!15]}]{@}{@}}
]
theorem sqrt2_not_rational: "&$\sqrt{2}$& &$\notin\;\mathbb{Q}$&"
  RULE proof_by_contradiction
  CONSIDER "&$\exists$&mn. |&$\sqrt{2}$&| = m/n &$\land$& coprime m n" END
  HAVE "m^2 = (&$\sqrt{2}$&)^2 * n^2" END
  HAVE eq: "m^2 = 2 * n^2" END
  HAVE "2 dvd m^2" END WITH eq
  HAVE "2 dvd m" END
  HAVE "2 dvd n"
    CONSIDER &\namedLabel{k}{\ensuremath{k}}{\texttt{k}}& where "m = 2 * k" END
    HAVE "2 * n^2 = 2^2 * k^2" END WITH eq
    HAVE "2 dvd n^2" END
  END
  HAVE "2 dvd gcd m n" END
  HAVE "2 dvd 1" END
END
\end{lstlisting}
\vspace{4em}
\caption{Minilang}
\label{fig:minilang-sqrt2}
\end{subfigure}
\normalsize
    \caption{
Minilang and mainstream proof languages on the same proof goal.
While Lean and Isar incorporate natural-language-like structures, (e.g., build-in keywords such as\texttt{have} and \texttt{hence}), their proofs are interspersed with tactics and other expert-oriented constructs that complicate learning.
In contrast, Minilang aims to highlight the key proof outlines with close alignment to informal proofs.
%Several proof languages. $(a \mathrel{|} b)$ and $(a \mathrel{\texttt{dvd}} b)$ denote integer division of $a$ by $b$.
    }
    \Description{}
    \label{fig:language}
\end{figure}

% 更详细的 Captions 哪个更好哪个更坏，为什么

\section{Motivation}\label{sec:motivation}

Early NTP works adopt LLM-guided proof search, where language models recommend the next tactic based on the current proof goal to heuristically guide the search process~\cite{BFS-Prover}.
As LLMs continue to advance in reasoning capabilities, recent works have shifted toward \emph{declarative theorem proving}~\cite{DPSKv2,Kimina,SeedProver,DSP}.
In this paradigm, language models construct proofs in a manner closer to informal pen-and-paper proofs --- by declaring a series of subgoals to plan and propose proof strategies from a high-level perspective --- thereby enabling models to more effectively leverage proof techniques learned from informal reasoning tasks.

In some purely declarative NTP works~\cite{Thor,DSP}, language models are used solely to draft high-level proof outlines, delegating all proof obligations to ATPs.
The rationale is that language models are better suited for planning high-level proof strategies rather than fine-grained reasoning, while proof obligations of subgoals are better handled by ATPs with their deterministic search algorithms.

Ideally, the training and generation of declarative NTP models should be based on a simple formal language that can succinctly express proof outlines while maintaining alignment with informal pen-and-paper proofs.
However, such an ideal language does not yet exist, despite the presence of all required declarative constructs in existing proof languages.

The problems are at least threefold: an overabundance of expert features\footnote{\xqym{Detailed descriptions and configuration options that provide fine-grained control over the strategy of reasoning tools.}}, extensive syntactic redundancy, and substantial requirements for proof automation in declarative proving.

These expert features include specialized mechanisms and tactics for fine-grained control over the reasoning process.
Such mechanisms include Isabelle's meta-logic encoding (e.g., generalized elimination), Rocq's \texttt{Ltac} tactic programming language, and Lean's \texttt{conv} mode for targeted rewriting.
Such tactics include Isabelle's \texttt{blast lim: $n$} for tableaux reasoning with specific search depth, Rocq's \texttt{rewrite} ... \texttt{at} ... for specifying rewrite locations, and Lean's \texttt{aesop} with customizable rule sets.
These features involve numerous configurations and technical details that impose substantial learning burdens on language models, particularly in NTP scenarios where training data is scarce.

Syntactic redundancy is particularly pronounced in Isabelle, the proof assistant underlying our work.
Isabelle's proof language \emph{Isar} admits multiple, often interchangeable proof idioms --- such as fact chaining via connectives versus labels, and structured proofs versus apply-style scripts --- yielding numerous ways to formalize identical proof procedures.
The syntactic redundancy addressed in this work includes at least: 4 mechanisms for opening proof contexts (\texttt{subgoal}, \texttt{proof}, \texttt{\{\}}, \texttt{goal\_cases}),
5 ways to apply tactics (\texttt{apply}, \texttt{by}, \texttt{proof}, \texttt{qed}, tactic combinators),
11 constructs for passing facts (\texttt{using}, \texttt{use-in}, \texttt{from}, \texttt{with}, \texttt{then}, \texttt{thus}, \texttt{hence}, \texttt{also}, \texttt{moreover}, \texttt{ultimately}, \texttt{finally}),
6 syntactic sugar for referring to facts and terms (\texttt{this}, \texttt{that}, \texttt{assms}, \texttt{prems}, \texttt{?thesis}, \texttt{?case}), and many semantically similar operations with subtle differences (\texttt{unfolding} vs. \texttt{unfold\_tac} vs. \texttt{subst}, \texttt{cases} vs. \texttt{case\_tac}, \texttt{induction} vs. \texttt{induct} vs. \texttt{induct\_tac}).

While redundant syntax may be harmless to humans, it becomes problematic for NTP given corpus scarcity: it either dilutes limited training data or forces models to consume substantial data to learn that different forms are semantically equivalent.

Finally, if NTP models are to focus on proposing high-level proof outlines, the underlying proof assistant must provide powerful ATPs to discharge the proof obligations of the subgoals raised in the proof outlines.
The capabilities of ATPs and NTP models are complementary: when ATP support is insufficient, NTP models must generate more fine-grained subgoals to make them amenable to automated proving.
In the worst case, some declarative NTP systems must handle proof obligation discharge themselves, even resorting to using additional NTP models in place of ATPs.

These three challenges motivate us to propose Minilang, a minimalist proof language, and Sledgehammer*, an enhanced version of the well-known ATP, Sledgehammer.
By \emph{minimalist}, we mean Minilang's design aims to minimize its language constructs, retaining only declarative structures and operations that align well with informal proofs.

%We elaborate on both contributions in the remainder of this paper.

% 概述目前主流的 declarative NTP 思路

% 展开 ITP 的 expert features

% 展开 Isabelle 的问题

%\input{sections/intro}
%\input{sections/isar}
\section{Minilang: A Minimalist Proof Language Mirroring Informal Proofs}\label{sec:minilang}

\begin{table}[t]
\centering
\caption{Core constructs of Minilang}
\label{tab:operations}
\resizebox{0.9\columnwidth}{!}{
\begin{tabular}{lp{11cm}}
\toprule \bf Operation & \bf Description \\\midrule
\multicolumn{2}{l}{\rule{0pt}{2ex}\small\it\color{gray!70!black} Declarative constructs}\\[.25em]
\minilang{HAVE} & decomposing a proof goal into step-by-step subgoals.\\
$\text{\minilang{CONSIDER}}_\lor$ & analyzing a proof goal by cases, e.g., consider the cases where x is positive, zero, or negative. \\
$\text{\minilang{CONSIDER}}_\exists$ & binding variables to the witnesses of existential statements, e.g., consider a number $p$ such that $p$ is a prime greater than $2025$.\\
\multicolumn{2}{l}{\rule{0pt}{2.5ex}\small\it\color{gray!70!black} Proof operations commonly found in informal proofs}\\[.25em]
\minilang{RULE} & proving a goal by a specific mode of argument, e.g., arguing by contradiction for a given goal, and deriving $A \longrightarrow B$ and $B \longrightarrow A$ to show $A \longleftrightarrow B$.\\
\minilang{CHOOSE} & proving an existential statement by providing a witness. \\
\minilang{SIMPLIFY} & equivalently rewriting the proof goal into a simpler form.\\
\minilang{CASE\_SPLIT} & applying structural case analysis to the goal.\\
\minilang{INDUCT} & applying induction to the goal.\\
%\minilang{DEFINE} & defining a local term \\
\minilang{END WITH} $\mathit{ps}$    & indicates that the target proof goal straightforwardly follows from the given premises $\mathit{ps}$.\\
\minilang{NEXT} & concludes the current goal's proof and moves to the next sibling goal when sibling goals exist. In terms of formalized semantics, it is an alias of \minilang{END}.  \\
\multicolumn{2}{l}{\rule{0pt}{2.5ex}\small\it\color{gray!70!black} Necessary technical command}\\[.25em]
\minilang{INTRO} & for management of variable and hypothesis context
\\\bottomrule
\end{tabular}}
\end{table}

%Minilang follows the purely declarative paradigm discussed in \cref{sec:motivation}, with emphasis on two guiding principles: minimalism and alignment with informal mathematical reasoning. \pw{Any details about these two principles?}

Minilang's constructs, as listed in \cref{tab:operations}, consist of (1) declarative constructs for decomposing hard proof goals into simpler subgoals, (2) operations commonly found in informal proofs, and (3) one necessary technical command.
These constructs, though minimalist, are sufficient for drafting high-level proof outlines, fulfilling the goals of the purely declarative paradigm.
%\pw{Why are these constructs minimal?}
%
Minilang's high-level design does not incorporate a tactic system, instead delegating all proof obligations to ATPs.

As a pragmatic decision to maximize the success of our Isar translation procedure, we additionally extend Minilang with a small number of escape hatches back to Isar, which are generated infrequently in the training corpus but allow us to avoid individual local translation failures invalidating multiple dependent definitions (\cref{step:translate}).

% Notably, Minilang by design does not provide a tactic system but instead delegates all proof obligations to ATPs.
% Our rationale, consistent with prior works~\cite{Thor,DSP}, is that language models are better suited for planning routes toward proofs from a high-level perspective, while concrete proof obligations can be delegated to specialized ATPs that excel at such tasks.

%This minimal set suffices to articulate high-level proof structures, while all detailed proof obligations are delegated to automated theorem provers.

This section is organized as follows:
\cref{sec:syntax} presents Minilang's syntax;
\cref{sec:proof-model} builds the proof model on which the semantics is formalized;
\cref{sec:semantics} elaborates on the semantics;
\cref{sec:sledgehammer} details Sledgehammer*, the ATP used in MiniLang;
\cref{sec:soundness} discusses MiniLang's soundness and relative completeness.

\subsection{Syntax}~\label{sec:syntax}

A Minilang proof script comprises a sequence of commands whose syntax is defined as follows.
The question mark (?) means that the clause can be omitted if the $\mathit{facts}$ is empty.
\small\begin{align*}
\text{Proof Script} \Coloneqq&~ \text{Command}^+\\[-.1em]
\text{Command} \Coloneqq&~ \text{\minilang{HAVE}}\ \mathit{props}
\mathrel{|} \text{\minilang{CONSIDER}}\ \mathit{props} \mathrel{|} 
\text{\minilang{INTRO}}
\\[-.3em]
|&~ \text{\minilang{RULE}}\ \mathit{fact}
\mathrel{|} \text{\minilang{CHOOSE}}\ \mathit{term}
\mathrel{|} \text{\minilang{SIMPLIFY}}\ \mathit{facts}^?\\[-.3em]
|&~ \text{\minilang{INDUCT}}\ \textit{the same argument syntax as Isabelle's \texttt{induct} tactic}\\[-.3em]
%\mathit{terms}\ (\text{\minilang{arbitrary:}}\ \mathit{vars})^?\ (\text{rule:}\ \mathit{fact})^?\\[-.3em]
|&~ \text{\minilang{CASE_SPLIT}}\ \textit{the same argument syntax as Isabelle's \texttt{cases} tactic}\\[-.3em]
%\mathit{terms}\ (\text{rule:}\ \mathit{fact})^?\\[-.3em]
|&~ \text{\minilang{END}}\ (\text{\minilang{WITH}}\ \mathit{facts})^?\ (\text{\minilang{WITHOUT}}\ \mathit{facts})^?\\[-.3em]
|&~ \text{\minilang{NEXT}}\ (\text{\minilang{WITH}}\ \mathit{facts})^?\ (\text{\minilang{WITHOUT}}\ \mathit{facts})^?
\end{align*}\normalsize
%These commands consist of all those listed in \cref{tab:operations} and three additional commands for either clarification purposes (\minilang{INTRO}) or syntax sugargs (\minilang{CASE_SPLIT} and \minilang{NEXT}).

\subsection{Proof Model: State Machine over Trees}~\label{sec:proof-model}

\input{figures/statemachine}
\begin{lrbox}{\codebox}
\begin{minipage}{6.5cm}
\begin{lstlisting}[language=MiniLang,
style=XXX,
moredelim={**[is][{\btHL[fill=cyan!15]}]{@}{@}},
numbers=left,
numberstyle=\footnotesize\color{gray}
]
theorem sqrt2_not_rational: "sqrt 2 &$\notin\;\mathbb{Q}$&"
  RULE proof_by_contradiction
  LET ?x = "sqrt 2"
  CONSIDER "&$\exists$&mn. |?x| = m / n &$\land$& coprime m n"&\label{line:mn}&
      END
  HAVE &\namedlabel{B}{\texttt{B}}&: "m^2 = ?x^2 * n^2" END
  HAVE eq: "m^2 = 2 * n^2" END
  HAVE &\namedlabel{C}{\texttt{C}}&: "2 dvd m^2" END WITH eq
  HAVE &\namedlabel{D}{\texttt{D}}&: "2 dvd m" END
  HAVE &\namedlabel{E}{\texttt{E}}&: "2 dvd n" &\label{line:E}&
    CONSIDER "&$\exists$\namedLabel{k}{\ensuremath{k}}{\texttt{k}}&. m = 2 * k" &\label{line:k}&
      END
    HAVE "2 * n^2 = 2^2 * k^2" END WITH eq
    HAVE "2 dvd n^2" END
  END
  HAVE I: "2 dvd gcd m n" END
  HAVE J: "2 dvd 1" END
END
\end{lstlisting}
\end{minipage}
\end{lrbox}

\begin{figure}[!tp]
\centering
\begin{minipage}{.46\textwidth}
\vspace*{3em}
\begin{tikzpicture}
\node(State)[draw,circle]{Tree};
\node(Init)[above=5mm of State, xshift=0.7cm]{%
\parbox{6cm}{\centering Initial State: $(\CCTX{\varnothing,\varnothing})\vdash\CGOAL{Goal}$\\ a tree with a single leaf}};
\node(Exit)[draw,below=18mm of State,inner sep=4pt]{%
\parbox{1.35cm}{\centering Proof\\\small
completes}};
\draw[-{Stealth[length=2.5mm]}] ($(Init.south)+(-0.7cm,0)$) -- (State);
\draw[->, >={Stealth[length=2.5mm]}] (State) edge[loop right] node {Command} ();
\draw[-{Stealth[length=2.5mm]}] (State) -- (Exit) node [midway, right, yshift=1mm] {\parbox{4cm}{
\minilang{END}\\
\small if the state is a leaf that is\\
\small provable by Sledgehammer*}};
% node [midway, right, yshift=-7mm] {\parbox{4cm}{
% if the state is a leaf that is\\
% provable by Sledgehammer*}};
\end{tikzpicture}
\captionof{figure}{Minilang's state machine.}
\label{fig:state-machine}

\vspace*{7em}

\begin{lstlisting}[language=MiniLang,
style=XXX,
moredelim={**[is][{\btHL[fill=cyan!15]}]{@}{@}},
numbers=left,
numberstyle=\footnotesize\color{gray},
numbersep=4pt,
xleftmargin=0.8em
]
theorem sqrt2_not_rational: "&$\sqrt{2} \notin\;\mathbb{Q}$&"
  RULE proof_by_contradiction
  CONSIDER "&$\exists$&mn. &$|\sqrt{2}| = \frac{m}{n} \land \mathrm{coprime}\,m\,n"$\label{line:mn}&
      END
  HAVE &\namedlabel{B}{\texttt{B}}&: "&$m^2 = (\sqrt{2})^2 \cdot n^2$&" END
  HAVE eq: "&$m^2 = 2 \cdot n^2$&" END
  HAVE &\namedlabel{C}{\texttt{C}}&: "&$2 \mathrel{\mathrm{dvd}} m^2$&" END WITH eq
  HAVE &\namedlabel{D}{\texttt{D}}&: "&$2 \mathrel{\mathrm{dvd}} m$&" END
  HAVE &\namedlabel{E}{\texttt{E}}&: "&$2 \mathrel{\mathrm{dvd}} n$&" &\label{line:E}&
    CONSIDER "&$\exists\namedLabel{k}{\ensuremath{k}}{\texttt{k}}.\; m = 2k$&" &\label{line:k}&
      END
    HAVE "&$2 * n^2 = 2^2 \cdot k^2$&" END WITH eq
    HAVE "&$2 \mathrel{\mathrm{dvd}} n^2$&" END
  END
  HAVE I: "&$2 \mathrel{\mathrm{dvd}} (\mathrm{gcd}\,m\,n)$&" END
  HAVE J: "&$2 \mathrel{\mathrm{dvd}} 1$&" END
END
\end{lstlisting}
\captionof{figure}{An example written in Minilang.}
\label{fig:sqrt2-example}
\end{minipage}\hspace*{.02\textwidth}%
\begin{minipage}{.52\textwidth}\small\centering
\begin{tikzpicture}
  %\node[draw, inner sep=2pt] (code) at (7.3,-3.5) {\usebox{\codebox}};
  
  \node (S1) {%
    $(\varnothing,\varnothing) \vdash \sqrt{2} \notin \mathbb{Q}$
    %\begin{forest} for tree={align=center, l sep=0, s sep=0.3cm}
    %  [$\cdots$
    %    [$\CCTX{(\Theta, \Gamma)}\vdash \CGOAL{G}$]
    %    [$\cdots$]
    %  ]
    %\end{forest}
  };

  \node (S2) [below=5mm of S1] {%
    $(\varnothing,\{\sqrt{2} \in \mathbb{Q}\}) \vdash \mathrm{False}$
};
  \draw[->,thick] ($(S1.south)$) -- ($(S2.north)$) node [midway, xshift=7mm] {\color{gray}line 2};
  
  \node (S4) [below=5mm of S2, inner sep=2pt] {%
    \begin{forest} for tree={align=center, l sep=0, s sep=0.3cm},
    tikz+={\node[anchor=north] at ([xshift=0mm,yshift=-9mm].south)
    {where $A = |\sqrt{2}| = \frac{m}{n} \land \mathrm{coprime}(m,n)$};},
      [$\CCTX{(\varnothing, \{\sqrt{2} \in \mathbb{Q}\})}$
        [$\CCTX{(\varnothing, \varnothing) \vdash \exists mn.\;A}$]
        [$\CCTX{(\{m,n\}, \{A\}) \vdash \mathrm{False}}$]
      ]
    \end{forest}
  };
  \draw[->,thick] ($(S2.south)$) -- ($(S4.north)$) node [midway, xshift=8.5mm] {\color{gray}line 3};

  \node (S5) [below=5mm of S4] {%
    $(\{m,n\}, \{\sqrt{2} \in \mathbb{Q}, A\}) \vdash \mathrm{False}$
  };
  \draw[->,thick] ($(S4.south)$) -- ($(S5.north)$) node [midway, xshift=7mm, color=gray] {line 4};
  
  \node (S6) [below=5mm of S5] {%
    $(\{m,n\}, \{\sqrt{2} \in \mathbb{Q}, A, B\}) \vdash \mathrm{False}$
  };
  \draw[->,thick] ($(S5.south)$) -- ($(S6.north)$) node [midway, xshift=7mm, color=gray] {line 5};
  
  \node (S9) [below=5mm of S6] {%
    $(\{m,n\}, \{\sqrt{2} \in \mathbb{Q}, A, B, eq, C, D\}) \vdash \mathrm{False}$
  };
  \node (S9') [below=0mm of S9, xshift=5mm, yshift=1mm] {
    \small where $B,eq,C,D$ are the propositions at line 5,6,7,8.
  };
  \draw[->,thick] ($(S6.south)$) -- ($(S9.north)$) 
  node [midway, xshift=8.5mm, color=gray] {line 6-8};
  
  \node (S10) [below=5mm of S9', xshift=-5mm] {%
    \begin{forest} for tree={align=center, l sep=0, s sep=0.3cm}
      [$\CCTX{(\{m,n\}, \{\sqrt{2} \in \mathbb{Q},\,A,B,eq,C,D\})}$
        [$\CCTX{(\varnothing, \varnothing) \vdash 2 \mathrel{\mathrm{dvd}} n}$]
        [$\CCTX{(\varnothing, \{2 \mathrel{\mathrm{dvd}} n\}) \vdash \mathrm{False}}$]
      ]
    \end{forest}
  };
  \draw[->,thick] ($(S9'.south)+(-5mm,0)$) -- ($(S10.north)$) node [midway, xshift=7mm, color=gray] {line 9};
  
  \node (S11) [below=5mm of S10, xshift=5mm, draw, dashed] {%
    \begin{forest} for tree={align=center, l sep=0, s sep=0.3cm}
      [$\CCTX{(\{m,n\}, \{\sqrt{2} \in \mathbb{Q},\,A,B,eq,C,D\})}$
        [$\CCTX{(\varnothing, \varnothing)}$
            [$\CCTX{(\varnothing, \varnothing) \vdash \exists k.\;m = 2k}$]
            [$\CCTX{({k}, {m = 2k}) \vdash 2 \mathrel{\mathrm{dvd}} n}$]]
        [$\CCTX{(\varnothing, \{2 \mathrel{\mathrm{dvd}} n\}) \vdash \mathrm{False}}$]
      ]
    \end{forest}
  };
  \draw[->,thick] ($(S10.south)$) -- ($(S11.north)+(-5mm,0)$) node [midway, xshift=7mm, color=gray] {line 10};

  \node (S12) [below=5mm of S11] {
    \begin{forest} for tree={align=center, l sep=0, s sep=0.3cm},
      [$\CCTX{(\{m,n\}, \{\sqrt{2} \in \mathbb{Q},\,A,B,eq,C,D\})}$
        [$\CCTX{(\{k\}, \{m = 2k\}) \vdash 2 \mathrel{\mathrm{dvd}} n}$]
        [$\CCTX{(\varnothing, \{2 \mathrel{\mathrm{dvd}} n\}) \vdash \mathrm{False}}$]
      ]
    \end{forest}
  };
  \draw[->,thick] ($(S11.south)+(-5mm, 0)$) -- ($(S12.north)+(-5mm,0)$) node [midway, xshift=7.5mm, color=gray] {line 11};

  \node (S17) [below=5mm of S12,xshift=-5mm] {
    $(\{m,n\}, \{\sqrt{2} \in \mathbb{Q},\,A,B,eq,C,D,E,I,J\}) \vdash \mathrm{False}$
  };
  \draw[->,thick] ($(S12.south)+(-5mm,0)$) -- ($(S17.north)$) node [midway, xshift=9.5mm, color=gray] {line 12-16};

  \node (S18) [below=5mm of S17, draw] {
    Proof completes
  };
  \draw[->,thick] ($(S17.south)$) -- ($(S18.north)$) node [midway, xshift=7.5mm, color=gray] {line 17};
\end{tikzpicture}
\captionof{figure}{The transition of the tree state in the example of \cref{fig:sqrt2-example}. Each node represents the state after the execution of the labeled line.}
\label{fig:long-example}
\end{minipage}
\Description{}
\end{figure}

The semantics of Minilang commands are defined as transitions over a state machine (see~\cref{fig:state-machine}).
An example is illustrated in~\cref{fig:long-example}.
%For example, the state transition of the \texttt{sqrt2\_not\_rational} example (see~\cref{fig:sqrt2-example}) is illustrated in \cref{fig:long-example}.
%Consider the example in~\cref{fig:sqrt2-example}, its state transition is illustrated in~\cref{fig:long-example}.
Specifically, a state is either the special proof completion state or a labelled tree that hierarchically organizes subgoals into contexts,
\small\begin{align*}
\textbf{State} &\Coloneqq \text{Tree} \qquad
\namedlabel{def:tree}{\text{Tree}} \Coloneqq \text{Leaf} \mathrel{|} ({\color{gray}\textit{label:}}\ \text{Context}, {\color{gray}\textit{children:}}\ \text{Tree}^+)
\\[-.1em]
\text{Context} &\Coloneqq (\text{a set of variables}, \text{a set of named hypotheses})\\[-.1em]
\text{Leaf} &\Coloneqq \text{Context} \vdash \text{Goal}
\qquad
\text{Goal} \Coloneqq \text{Term}
\end{align*}\normalsize
where leaves represent unproven subgoals, and internal nodes group sibling subgoals that inherit a common context from decomposing a larger goal (e.g., lines \ref{line:mn}, \ref{line:E}, \ref{line:k}).

For example, the resulting tree after executing the line \ref{line:k} of \cref{fig:sqrt2-example} (\minilang{CONSIDER} $\exists k.\;m = 2 * k$) is 
given in the dashed box of \cref{fig:long-example}.
% \begin{center}
% \begin{forest} for tree={align=center, l sep=-0.1cm, s sep=0.3cm}
%   [$\CCTX{(\{m,n\},\{\sqrt 2 \in \mathbb{Q}, \mathrm{A}, \ref{B}, \ref{C}, \ref{D}\})}$
%     [$\CCTX{(\varnothing,\varnothing)}$
% [$\CCTX{(\varnothing,\varnothing)\vdash \exists k.\,m = 2k}$]
% [$\CCTX{(\{\ref{k}\},\{m = 2k\})} \vdash\CGOAL{2\mathrel{\mathrm{dvd}} m}$]]
%     [{$\CCTX{(\varnothing, \{2 \mathrel{\mathrm{dvd}} m\})} \vdash\CGOAL{\mathrm{False}}$}]
%   ]
% \end{forest}
% \end{center}
This tree contains three subgoals arranged left to right. The leftmost subgoal (opened by line \ref{line:k}) requires proving the existence of $k$.
The middle subgoal, opened by \minilang{HAVE E} at line \ref{line:E}, is from the parent of the first subgoal. Its context has variable $k$ fixed with condition $m = 2k$, allowing the proof for the middle subgoal to use this condition once the first subgoal establishes $k$'s existence.
The rightmost subgoal (False) is the top goal of the entire proof. It can similarly utilize the conclusion $(2 \mathrel{\mathrm{dvd}} m)$ from the middle subgoal. Each leaf subgoal's context includes all labeled contexts from its ancestors in the tree, so all subgoals can access the previously established lemmas A, \ref{B}, \ref{C}, \ref{D}.

Given the top-level proof goal $G$, the \emph{initial state} is the tree with a single root node $((\varnothing,\varnothing) \vdash G)$ that represents the top goal $G$ itself. In the \texttt{sqrt2\_not\_rational} example, this initial state is,
\begin{equation}
\left((\varnothing,\varnothing) \vdash \sqrt{2} \notin \mathbb{Q}\right) 
\label{node:root}
\tag{Initial State}
\end{equation}

Given a proof script as a sequence of commands, Minilang's proof system verifies this proof script by executing each of its commands successively to transition the state machine. This execution yields three results:
\begin{enumerate}
\item The execution gets stuck at one of the commands because no corresponding transition rule can be found, indicating proof failure.
\item All commands are successfully executed through to the last one, and the machine reaches a special terminal state (the square node in \cref{fig:state-machine}) that represents proof completion.
\item All commands are successfully executed through to the last one, but the machine does not reach the proof-complete state, indicating that the proof script is incomplete and the proof remains unfinished.
\end{enumerate}

\vspace{-1em}
\subsection{Semantics}\label{sec:semantics}

\input{figures/semantics}

\newcommand{\XRS}{\tikz[baseline=-0.5ex, x=1ex, y=1ex] \node {\begin{forest} for tree={align=center, l sep=-100mm, s sep=0.8cm, inner sep=0pt, outer sep=1.5pt, scale=0.85}, for level={1}{yshift=8mm}
      [$\mathcal{R}$
        [$X$]
        [$\mathcal{S}$]
      ]%
\end{forest}};}

The semantics of Minilang commands are formalized as transition schemas presented in \cref{fig:semantics}.
Each transition rewrites only the leftmost leaf or the leftmost non-leaf node of the state tree; consequently, it suffices to describe the change at that leftmost position.
We use a schematic diagram \XRS to represent an arbitrary tree whose leftmost leaf is $X$. The schematic variable $\mathcal{R}$ represents $X$'s parent node and all its upward context, while the schematic variable $\mathcal{S}$ represents potentially multiple sibling subtrees of $X$.
As a special case, a schematic diagram \XRS can be instantiated to a tree with a single root node $X$, where $\mathcal{R}$ and $\mathbf{S}$ are considered instantiated to empty.

Since all transitions act on the leftmost subgoal, we refer to that leftmost goal as the current goal in what follows.
We now describe each command’s semantics.

\minilang{HAVE} $G_1,\cdots,G_n$ is the key to decomposing a proof goal $G_0$ into subgoals $G_1,\cdots,G_n$.
It decomposes the proof problem about the goal $G_0$ into: 1) proving the subgoals $G_1,\cdots,G_n$ successively; and then 2) using their proved conclusions as lemmas to prove the original goal $G_0$.

\begin{figure}[t]
\centering
\begin{subfigure}[b]{0.53\linewidth}
\begin{tikzpicture}[scale=0.9, transform shape]
  \node (LEFT) {%
    \begin{forest} for tree={align=center, l sep=0.15cm, s sep=1mm}
      [$\mathcal{R}$
        [$\CCTX{(\Theta, \Gamma)}\vdash \CGOAL{G}$]
        [$\mathcal{S}$]
      ]
    \end{forest}
  };
  
  \node (RIGHT) [right=-20mm of LEFT,yshift=-5mm] {%
    \begin{forest} for tree={align=center, l sep=0.15cm, s sep=1mm}
      [$\mathcal{R}$
        [$\CCTX{(\Theta,\Gamma)}$
          [$\CCTX{(\varnothing,\varnothing)}\vdash\NOTE{P_1 \lor \cdots \lor P_n}$]
          [$\CCTX{(\varnothing,\NOTE{\{P_1\}})}\vdash \CGOAL{G}$]
          [$\cdots$]
          [$\CCTX{(\varnothing,\NOTE{\{P_n\}})}\vdash \CGOAL{G}$]
        ]
        [$\mathcal{S}$]
      ]
    \end{forest}
  };
  
  \draw[->, thick] ($(LEFT.east)+(-5mm,3mm)$) -- ($(RIGHT.west)+(50mm,8mm)$) node[midway, above] {\small\minilang{CONSIDER} $P_1$ | $\cdots$ | $P_n$} ;
\end{tikzpicture}
\caption{\small One function: Case-analysis}
\label{fig:consider-cases}
\end{subfigure}%
\begin{subfigure}[b]{0.47\linewidth}
\begin{tikzpicture}[scale=0.9, transform shape]
  \node (LEFT) {%
    \begin{forest} for tree={align=center, l sep=0, s sep=0}
      [$\mathcal{R}$
        [$\CCTX{(\Theta, \Gamma)}\vdash \CGOAL{G}$]
        [$\mathcal{S}$]
      ]
    \end{forest}
  };
  
  \node (RIGHT) [right=-5mm of LEFT, yshift=-5mm] {%
    \begin{forest} for tree={align=center, l sep=0.15cm, s sep=1mm}
      [$\mathcal{R}$
        [$\CCTX{(\Theta,\Gamma)}$
          [$\CCTX{(\varnothing,\varnothing)}\vdash\NOTE{\exists \bar{x}.\;P(\bar{x})}$]
          [$\CCTX{(\bar{x},\NOTE{\{P(\bar{x})\}})}\vdash \CGOAL{G}$]
        ]
        [$\mathcal{S}$]
      ]
    \end{forest}
  };
  
  \draw[->, thick] ($(LEFT.east)+(-5mm,3mm)$) -- ($(RIGHT.west)+(30mm,8mm)$) node[midway, above] {\small\minilang{CONSIDER} $\exists\bar{x}.\;P(x)$};
\end{tikzpicture}
\caption{\small The other function: Extential-Witness Extraction}
\label{fig:consider-ex}
\end{subfigure}
\caption{The semantics of \minilang{CONSIDER} combines two functions.}
\Description{}
\label{fig:decomposed-consider}
\end{figure}

\minilang{CONSIDER} combines two functions: (a) case-analysis (e.g., consider the cases where x is positive, zero, or negative) and (b) existential-witness extraction (e.g., let $p$ be a prime greater than $2025$).
This design is logically reasonable because both of the functions perform elimination of disjunctive connectives ($\exists$ and $\lor$).
However, for clarity, we elaborate on these two functions separately.

Illustrated in \cref{fig:consider-cases}, $\text{\minilang{CONSIDER}}\ P_1 \mathrel{|} \cdots \mathrel{|} P_n$ splits the proof of the current goal $G$ into $n$ cases.
%\begin{center}
%\begin{tikzpicture}
%  \node (LEFT) {%
%    \begin{forest} for tree={align=center, l sep=0.15cm, s sep=1mm}
%      [$\mathcal{R}$
%        [$\CCTX{(\Theta, \Gamma)}\vdash \CGOAL{G}$]
%        [$\mathcal{S}$]
%      ]
%    \end{forest}
%  };
%  
%  \node (RIGHT) [right=-2mm of LEFT,yshift=-5mm] {%
%    \begin{forest} for tree={align=center, l sep=0.15cm, s sep=1mm}
%      [$\mathcal{R}$
%        [$\CCTX{(\Theta,\Gamma)}$
%          [$\CCTX{(\varnothing,\varnothing)}\vdash\NOTE{P_1 \lor \cdots \lor P_n}$]
%          [$\CCTX{(\varnothing,\NOTE{\{P_1\}})}\vdash \CGOAL{G}$]
%          [$\cdots$]
%          [$\CCTX{(\varnothing,\NOTE{\{P_n\}})}\vdash \CGOAL{G}$]
%        ]
%        [$\mathcal{S}$]
%      ]
%    \end{forest}
%  };
%  
%  \draw[->, thick] ($(LEFT.east)+(-3mm,0)$) -- ($(RIGHT.west)+(40mm,5mm)$) node[midway, above] {\small\minilang{CONSIDER} $P_1$ | $\cdots$ | $P_n$} ;
%\end{tikzpicture}
%\end{center}
This operation produces $n+1$ subgoals, where the first goal $P_1 \lor \cdots \lor P_n$ verifies that the case split is exhaustive, namely, $P_1,\cdots,P_n$ are all the possible situations and no other case is missing from consideration.
Once the exhaustiveness is proven, the proof proceeds with $n$ separate branches, each proving the goal under the assumption of case $P_i$.

Illusrated in \cref{fig:consider-ex}, \minilang{CONSIDER} $\exists \bar{x}.\;P(\bar{x})$ introduce a sequence of fresh variables $\bar{x}$ and bind them to certain terms satisfying the condition $P(\bar{x})$, once the existence of such terms is shown.
Regarding the notation, $\bar{x}$ represents a sequence of variables.
% \begin{center}
% \begin{tikzpicture}
%   \node (LEFT) {%
%     \begin{forest} for tree={align=center, l sep=0, s sep=0}
%       [$\mathcal{R}$
%         [$\CCTX{(\Theta, \Gamma)}\vdash \CGOAL{G}$]
%         [$\mathcal{S}$]
%       ]
%     \end{forest}
%   };
%   
%   \node (RIGHT) [right=10mm of LEFT] {%
%     \begin{forest} for tree={align=center, l sep=0.15cm, s sep=2mm}
%       [$\mathcal{R}$
%         [$\CCTX{(\Theta,\Gamma)}$
%           [$\CCTX{(\varnothing,\varnothing)}\vdash\NOTE{\exists \bar{x}.\;P(\bar{x})}$]
%           [$\CCTX{(\bar{x},\NOTE{\{P(\bar{x})\}})}\vdash \CGOAL{G}$]
%         ]
%         [$\mathcal{S}$]
%       ]
%     \end{forest}
%   };
%   
%   \draw[->, thick] ($(LEFT.east)+(-3mm,0)$) -- ($(RIGHT.west)+(22mm,0)$) node[midway, above] {\small\minilang{CONSIDER} $\exists\bar{x}.\;P(x)$};
% \end{tikzpicture}
% \end{center}
This command generates 2 subgoals: The first subgoal asserts the existence of such terms satisfying $P$; the second subgoal augments the original goal $G$ by binding $\bar{x}$ to the terms satisfying $P$.
%Finally, the complete semantics of \minilang{CONSIDER} is the combination of the two functions above, and is formalized in \cref{fig:semantics}.

HAVE and CONSIDER introduce lemmas within the current goal's context. However, we have not yet provided a command for introducing variables and hypotheses into the context. Without it, every context would remain trivially empty.
This is the motivation for the \minilang{INTRO} command.
The command rewrites the current goal by moving into the context all universally quantified variables ($\bar{x}$ in \cref{fig:semantics}) and all hypotheses ($\{H_i\}_{1\leq i \leq n}$ in \cref{fig:semantics}) occurring in the goal proposition, so that these variables and hypotheses are in scope for subsequent lemmas.

While \minilang{INTRO} is admittedly technical, it is necessary for explicit context management. Minilang offers an option to automatically invoke INTRO before HAVE and CONSIDER, allowing proofs to more closely resemble informal mathematics at the cost of less fine-grained control. In our experiments, we disable this option and require models to use INTRO explicitly.

% Although \minilang{INTRO} is admittedly technical and not fully aligned with 
% natural language-based informal proofs, it is necessary to make context management explicit.
% As a partial remedy, Minilang provides an option to automatically invoke INTRO before calling HAVE and CONSIDER. This is sometimes effective to let users avoid writing INTRO explicitly and yield proofs that more closely resemble informal mathematics, though at the cost of less fine-grained control over the context.
% In our experiments, we do not turn on this option and ask language models to learn and use this command.

\begin{figure}[t]
\centering
\vspace*{-1em}
\begin{tikzpicture}
  \node (LEFT) {%
    \begin{forest} for tree={align=center, l sep=0, l=9mm, s sep=0mm}
      [$\mathcal{R}$
        [$\CCTX{(\Theta_1, \Gamma_1)}$
        [$\CCTX{(\Theta_2, \Gamma_2)}\vdash G$]]
      ]
    \end{forest}
  };
  
  \node (RIGHT) [right=1mm of LEFT] {%
    \begin{forest} for tree={align=center, l sep=0, s sep=0mm}
      [$\mathcal{R}$
        [$\CCTX{(\Theta_1 \cup \Theta_2\,, \Gamma_1 \cup \Gamma_2)} \vdash G$ ]
      ] \end{forest}
  };
  
  \draw[->, thick] ($(LEFT.east)+(-3mm,0)$) -- ($(RIGHT.west)+(15mm,0)$)
  node[midway, above] {Reduction};

  \node (LEFT') [right=5mm of RIGHT] {%
    \begin{forest} for tree={align=center, l sep=0, l=8mm, s sep=0mm}
      [$\mathcal{R}$
        [$\CCTX{(\Theta_1, \Gamma_1)}$
        [$\CCTX{(\Theta_2, \Gamma_2)}$
         [$\mathcal{T}_1$] [$\cdots$] [$\mathcal{T}_n$]]]
      ]
    \end{forest}
  };
  
  \node (RIGHT') [right=15mm of LEFT'] {%
    \begin{forest} for tree={align=center, l sep=0, s sep=0mm}
      [$\mathcal{R}$
        [$\CCTX{(\Theta_1 \cup \Theta_2\,, \Gamma_1 \cup \Gamma_2)}$
            [$\mathcal{T}_1$] [$\cdots$] [$\mathcal{T}_n$]]
      ] \end{forest}
  };
  
  \draw[->, thick] ($(LEFT'.east)+(-5mm,0)$) -- ($(RIGHT'.west)+(0mm,0)$)
  node[midway, above] {Reduction};
\end{tikzpicture}
\caption{Minilang's semantics additionally incorporates two automatic reduction rules. $\mathcal{T}_1,\cdots,\mathcal{T}_n$ are schematic variables denoting arbitrary subtrees.}
\Description{}
\label{fig:autored}
\end{figure}

\minilang{END} and \minilang{NEXT} are syntactic aliases of one another that are semantically equivalent.
They correspond to the common idiom in informal proofs --- ``the goal is easy to show'' or ``the goal follows straightforwardly from the premises''.
Operationally, these commands conclude the proof of a subgoal by invoking Sledgehammer* to discharge its proof obligations.
If Sledgehammer* fails to prove the goal or times out, the proof execution terminates with an error.

The motivation of introducing the \minilang{NEXT} alias is solely for readability: the word ``next'' naturally reads as ``finish this subgoal and move to its sibling''.
%which makes it well suited for examples like the \cref{fig:consider-example}.

The first \minilang{END} rule in \cref{fig:semantics} removes the leftmost subgoal when it has siblings, but no rule handles the case where it is an only child. This does not compromise completeness: two automatic reduction rules (\cref{fig:autored}) maintain the invariant that every non-leaf node has at least two children. These rules collapse any single-child node into its parent by merging their contexts and are applied preemptively before all other rules.

% The first \minilang{END} rule in \cref{fig:semantics} specifies how the commands remove the leftmost goal when it has at least one sibling; however, no rule is given for the case where the leftmost goal is the only child of its parent.
% This does not render the reasoning system incomplete, because the semantics include two automatic reduction rules (\cref{fig:autored}) that maintain an invariant that every non-leaf node must have at least two children.
% These rules eliminate any non-leaf node with a single child by collapsing the internal node and its only child into a node whose context is the concatenation of the parent’s and the child’s contexts.
% The rules are applied automatically whenever possible and preemptively, ahead of all other rules, thereby maintaining the invariant.

The second \minilang{END} rule concludes the entire proof, transitioning to the terminal ``proof complete'' state. The rule applies only when the proof state is a single-node tree.

% Additionally, another use of \minilang{END} is to conclude the entire proof and transition the state machine to the terminal ‘proof complete’ state.
% This behavior is formalized by the second \minilang{END} rule in \cref{fig:semantics}. The rule applies only when the proof state is a one-node tree (a single root) --- the entire proof state contains exactly one remaining goal --- which is then discharged by Sledgehammer* (\cref{sec:sledgehammer}).

The foregoing commands comprise all declarative structures in Minilang. We now present commands that align with common proof operations in informal mathematics.

The \minilang{RULE} command corresponds to adopting a specific route of argument in an informal proof (e.g., proof by contradiction).
Such a route of argument is encoded as an inference rule of shape,
$H_1\longrightarrow \cdots \longrightarrow H_n \longrightarrow G$, 
which indicates that the goal $G$ can be reduced to the subgoals $\{H_i\}_{1 \leq i \leq n}$.
For example, the rule \texttt{proof\_by\_contradiction} in \cref{fig:sqrt2-example} has form, $(P \longrightarrow \mathrm{False}) \longrightarrow \lnot P$.

% For example, in the \texttt{sqrt2\_not\_rational} example (\cref{fig:sqrt2-example}), rule \texttt{proof\_by\_contradiction} instructs Minilang to argue by contradiction, reducing the goal $\lnot (\sqrt{2} \in \mathbb{Q})$ to $(\sqrt{2} \in \mathbb{Q} \longrightarrow \mathrm{False})$, that is, derive a contradiction from the assumption $\sqrt{2} \in \mathbb{Q}$.
% \[
% \text{for any $P$,}\qquad(P \longrightarrow \mathrm{False}) \longrightarrow \lnot P
% \tag{\,\texttt{proof\_by\_contradiction}\,}
% \]

%The rule $r$ may be specified explicitly in the proof script; otherwise, Minilang uses the default inference rule supplied by the Isabelle system via pattern matching over the goal's expression.
% For the goal $\lnot (\sqrt{2} \in \mathbb{Q})$, the default rule is precisely \texttt{proof\_by\_contradiction}.
% Hence, in the example of \cref{fig:language}, RULE \texttt{proof\_by\_contradiction} can be abbreviated to RULE.

The command \minilang{CHOOSE} supplies a witness for an existential goal. When the goal is to prove $\exists x.\;G(x)$, CHOOSE $y$ provides a concrete term $y$ and reduces the goal to proving $G(y)$.
The command \minilang{SIMPLIFY} $\mathit{args}$ corresponds to the simplification operation in informal proofs. It is implemented by invoking Isabelle’s system simplifier on the proof goal, with argument $\mathit{args}$. 
The command \minilang{CASE\_SPLIT} and \minilang{INDUCT} perform structural case analysis and induction.
They are implemented by Isabelle's \texttt{cases} and \texttt{induct} tactics.

\subsection{Sledgehammer*: An Improved Sledgehammer}\label{sec:sledgehammer}

All proof obligations in Minilang are delegated to an improved Sledgehammer which we call Sledgehammer*. The improvement (of Sledgehammer* itself and the way we incorporate it) includes three aspects.

\subsubsection{Premise selection aided by proof generation models}

%Sledgehammer consists of two components: a premise selection mechanism that identifies from the system database the relevant lemmas that may help to prove a given goal, and backend SMT solvers that use these selected lemmas to prove the goal.
%

The \minilang{END} and \minilang{NEXT} commands incorporate a specific syntax to annotate the relevant lemmas that Sledgehammer* could consider using as premises when carrying out the proof search, and the lemmas that should be avoided.
\[
\text{\minilang{END}\,/\,\minilang{NEXT WITH} \textit{relevant-lemmas} \minilang{WITHOUT} \textit{avoided-lemmas}}
\]
Our translation pipeline employs a specific pass (\cref{step:refinement}) to extract the relevant and avoided lemmas from ATP proofs and annotate them in the training corpus.
This allows proof generation models to learn the relevant and avoided lemmas through the annotations in the corpus, as well as how these lemmas relate to both the local proof step and the overall proof strategy.
Consequently, the models can generate relevant/avoided lemma annotations from a global proof strategy perspective.

On the implementation side, we reuse Sledgehammer's existing interface for relevant/avoided lemma hints.
Sledgehammer's original premise selection is based on heuristics~\cite{MePo} and classical machine learning methods like $k$-NN~\cite{MaSh}.
We continue to use this premise selection mechanism in addition to our explicit hints.
We include a preprocessing pass that filters out undefined references produced by language-model hallucinations. This prevents syntax errors and lets proofs proceed even when some annotations are invalid.

\subsubsection{Integration with simplification system}

The next improvement of Sledgehammer*is its integration with Isabelle's simplification system, a term rewriting engine enhanced with programmable simplification procedures.
This system is notably powerful because any decision problem in Isabelle/HOL can reduce to rewriting propositions to True, allowing arbitrary decision procedures to be embedded as simplification procedures.
These embedded procedures are often complementary to SMT-based Sledgehammer, and can discharge many goals that Sledgehammer alone fails to prove.

Specifically, the improvement comprises two parts:
First, tactic \texttt{auto} is applied before invoking Sledgehammer.
It rewrites the goal using the system simplifier and may split it into multiple, simpler subgoals --- cases where Sledgehammer often fails on the original goal but can succeed on the decomposed subgoals one by one.
However, \texttt{auto} can be time-consuming on complex goals, so we impose a timeout and, if it expires, fall back to the safer tactic \texttt{clarsimp}.
Second, Sledgehammer* runs the simplification-based brute-force tactic \texttt{fastforce} in parallel with Sledgehammer as an extra ATP backend.

Moreover, the relevant/avoided lemma hints are also used to augment the simplifiers invoked by Sledgehammer*.
A challenge is that these lemmas should not be fed naively as undifferentiated inference rules, because the simplifiers distinguish rule roles (e.g., rewriting, splitting, and introduction/elimination in tableaux-style reasoning).
We develop a heuristic to guess the intended role of a given rule and configure the simplifiers accordingly.

This heuristic combines name-based and expression-based analysis.
Rule names in Isabelle follow certain naming conventions --- for example, rewriting rules typically end in \texttt{\_simp}, and splitting rules typically end in \texttt{.splits}. If a rule's name matches one of these conventions, the heuristic classifies it accordingly.
Second, many role-specific rules exhibit characteristic shapes; for example, elimination rules always have the form $(\mathit{term}_1 \longrightarrow ?C) \longrightarrow \cdots \longrightarrow (\mathit{term}_n \longrightarrow ?C) \longrightarrow ?C$ where $?C$ is a schematic variable.
The heuristic pattern-matches such shapes in a rule’s statement to determine its role.
Finally, if none of these signals determines a role, the heuristic defaults to classifying the rule as a rewriting rule, which is the most common case.

\subsubsection{Caching proofs}
Another minor improvement is that Sledgehammer* records all obtained proofs in a persistent cache keyed by the goal’s expression. Because Sledgehammer is time-consuming and somewhat nondeterministic, caching results substantially accelerates proof replay.

\subsection{Soundness and Relative Completeness}\label{sec:soundness}
\begin{theorem}[Soundness]
If Isabelle's simplifiers, tactic \texttt{cases}, tactic \texttt{induct}, and Sledgehammer* are sound, then $\mathrm{Minilang}$ is sound: 
The execution of a proof script reaches the terminal ``proof completes'' state only if the target proof goal is semantically valid.
\end{theorem}

The formalization of Minilang's semantics is tied to Sledgehammer*.
If we generalize the semantics to be parametric in the choice of ATP (rather than fixed to Sledgehammer*), we can show that there exists an ATP with respect to which Minilang is at least as proof-theoretically complete as a Natural Deduction system.

\begin{theorem}[Relative Completeness]\label{thm:completeness}
Given a Natural Deduction system $\mathit{ND}$, there exists an $\mathit{ATP}$ such that any proposition provable by $\mathit{ND}$ is provable by Minilang with respect to the $\mathit{ATP}$.
\end{theorem}
\begin{proof}
The idea is to use HAVE statements to replay the proof tree of this provable proposition.
Write all the formulas in the tree's nodes into a sequence of HAVE statements, from the leaves to the root, by a post-order traversal.
Note that the proof for every statement requires only one \textit{ND} rule application.
We show that an ATP exists to prove all the HAVE statements:
Because the number of \textit{ND} rules and possible choices of their operands is finite, an ATP can enumerate all possible applications to find the correct one.
\end{proof}

%Operationally, \minilang{RULE} closely mirrors Isabelle’s \texttt{rule} tactic. All introduction rules in Isabelle are usable with \minilang{RULE}, allowing users to reason in Minilang directly with Isabelle’s rules.

% This does not make our reasoning system incomplete because we have an automatic reduction rule that ensures any 
% 
% demonstrate that the commands remove the leftmost goal if it has at least one sibling; however, the rule is missing for the case when the leftmost goal is the only child of its parent.
% When the leftmost 

%HAVE and CONSIDER provide sufficient constructs for declarative proofs, where %complex proof goals are recursively decomposed into subgoals simple enough for an $\mathit{ATP}$ to handle.
%These subgoals are then closed by END and NEXT without requiring further specification.

%For declarative proofs, HAVE and CONSIDER provide sufficient constructs for decomposing complex proof goals into subgoals simple enough for an $\mathit{ATP}$ to handle.
%These subgoals are closed by END and NEXT without requiring further specification.

%\newpage
%\input{sections/minilang}
\section{Translation from Isar to Minilang}\label{sec:translation}

Training NTP models requires substantial proof data, which does not exist for our newly designed language.
We address this through automated translation from Isar to Minilang, successfully converting 85.28\% of 340K existing Isar proofs obtained from Isabelle’s Archive of Formal Proofs (AFP)~\cite{AFP} and Isabelle/HOL system library.
%obtaining $\sim$285K proofs from AFP's $\sim$324K proofs.
%
%Moreover, this translation process highlights three key strategies that embody our minimalist idea for redesigning a proof language to enhance the performance of NTP:
%\pw{Where is the minimalist principle?}
% This translation process employs three strategies:
% \begin{itemize}
% \item \textbf{Elaboration}: Make implicit information explicit by exposing hidden details in a clear, structured manner;
% \item \textbf{Normalization}: Consolidate diverse approaches for achieving the same logical purpose;
% \item \textbf{Elimination of tactics}: Replace tactics with Sledgehammer*, except those that correspond well to informal proofs.
% \end{itemize}
% %
This process of eliminating non-essential components and concepts from Isar to ultimately obtain Minilang also reflects the differences between Minilang's minimalist design and Isar's design.
%Additionally, Minilang also encapsulates the meta layer of Isabelle/Pure, making it transparent.

%In what follows, \cref{sec:Isar} introduces the background of Isar; \cref{sec:sec:translation-detail} presents our translation pipeline; \cref{sec:translation-result} analyzes the results

\subsection{Background}\label{sec:Isar}
%\conrad{reframe to be purely background on Isar - 4.2 is where you start to talk about how you'll address it}

Isar is a powerful and intricate language. Its diverse syntax and rich features pose substantial challenges for translating into Minilang.
%This section provides background and describes some of the challenges, while deferring others to \cref{sec:normalization}, where we address them in depth.

\subsubsection{Declarative Proof Structure}\label{sec:isar-proof-block}

Isar provides a suite of declarative statements for various functions, such as \texttt{have} for introducing subgoals, \texttt{obtain} for extracting witnesses from existential statements, and \texttt{assume} for assumption management. All such statements must appear inside a declarative proof block.

\begin{figure}[t]
\centering
\begin{minipage}[t]{0.5\textwidth}
  \begin{lstlisting}[language=isar, style=XXX, basicstyle=\ttfamily\small]
&\it assume we face multiple proof goals $G_1,\cdots,G_4$&
proof (&\it some tactic&)
  have &\it some lemma $A$& by ...
  then show &$G_3$& by &\it a tactic that may use $A$&
  moreover with &$A$& have &\it some lemma $B$& by ...
  finally show &$G_1$& by &\it a tactic that may use $A,B,G_3$&
next
  obtain &$x$& where &\it $x$ satisfies some condition $C(x)$&
  hence &$G_2$& and &$G_4$& by &\it a tactic that may use $C(x)$&
qed
\end{lstlisting}
\captionof{figure}{An example of complications in Isar’s proof structure. \isar{hence} $\equiv$ \isar{then show} is an abbreviation.}
\label{fig:messy-isar}
\end{minipage}\hspace*{.025\textwidth}%
\rule[-4.5cm]{0.5pt}{4cm}%
\hspace*{.025\textwidth}%
\begin{minipage}[t]{0.45\textwidth}
\begin{lstlisting}[language=isar, style=XXX, basicstyle=\ttfamily\small]

proof (&\it some tactic&)
  have &$A$& by ...
  show &$G_3$& using &$A$& by &\it tactic&
  have &$B$& using &$A$& by ...
  show &$G_1$& using &$G_3,B$& by &\it tactic&
next
  obtain &$x$& where &$C(x)$&
  show &$G_2$& and &$G_4$& using &$C(x)$& by &\it tactic&
qed
\end{lstlisting}
\captionof{figure}{Normalization of the connectives in \cref{fig:messy-isar}.}
\label{fig:no-connective}
\end{minipage}
\Description{}
\end{figure}

% \begin{wrapfigure}{r}{0.5\textwidth}
%   \begin{lstlisting}[language=isar, style=XXX, basicstyle=\ttfamily\small]
% &\it assume we face multiple proof goals $G_1,\cdots,G_4$&
% proof (&\it some tactic&)
%   have &\it some lemma $A$& by ...
%   then show &$G_3$& by &\it a tactic that may use $A$&
%   moreover with &$A$& have &\it some lemma $B$& by ...
%   finally show &$G_1$& by &\it a tactic that may use $A,B,G_3$&
% next
%   obtain &$x$& where &\it $x$ satisfies some condition $C(x)$&
%   hence &$G_2$& and &$G_4$& by &\it a tactic that may use $C(x)$&
% qed
% \end{lstlisting}
%   \caption{An example of complications in Isar’s proof structure. \isar{hence} $\equiv$ \isar{then show} is an abbreviation.}
%   \label{fig:messy-isar}
% \end{wrapfigure}
% 
% \begin{wrapfigure}{r}{0.4\textwidth}
% \end{wrapfigure}
% 

A declarative proof block begins with \isar{proof} and ends with \isar{qed}.
When multiple goals are present, command \isar{next} can be inserted within a proof-qed block to close the current block and open a fresh one at the same level.
However, a complication is that, \isar{next}-separated blocks do not correspond one-to-one, in sequence, to the proof goals.
Instead, Isar uses statement \isar{show} \textit{goal} \isar{by} \textit{tactic} to explicitly target a particular \textit{goal} and discharge it with a \textit{tactic}. Moreover, a block may contain multiple \isar{show} statements proving multiple goals in any order, and even a single \isar{show} can target several goals. \cref{fig:messy-isar} illustrates an example.

This intricate proof context in Isar complicates translation to Minilang, because in Minilang proof blocks have a simple one-to-one, sequential correspondence with proof goals; the translation must therefore normalize Isar proofs to Minilang's block organization.

% Isar 提供一系列 declarative statement，例如 \isar{have}, \isar{obtain}, \isar{assume} 等用于引入 subgoals, obtaining witness of existential statemetns, and 控制 context
% Declarative Isar proof 通常由 \isar{proof} 命令导引开始，由 \isar{qed} 结束。\isar{proof} opens a context allowing declarations of proof operations for proving proof goals; \isar{qed} closes the context.
% 在有多个 proof goals 存在时，可以在 \isar{proof}-\isar{qed} block 中插入 \isar{next} 指令，闭合并重新开启一个 fresh proof context 以证明 next proof goals
% 然而 \isar{next}-separated sub-blocks 并不按序一一对应 the proof goals。
% 实际上 Isar 使用 \isar{show} a goal by tactic 来显示地作用一个 tactic 到一个 proof goal 上以对 this goal 进行证明。
% 糟糕的是，一个 block/sub-block 可以包含多个 \isar{show} statement 以对多个 proof goals 以任意允许进行证明；甚至一个 \isar{show} statement 也可以作用 tactics 到多个 proof goals 上。
% Figure 10 is such an example.
% Isar 的这个复杂 proof context 结构带给到 Minilang 的翻译诸多困难，因为 Minilang 中 declarative operations 与 proof goals 直接是简单按序一一对应的。

\subsubsection{Connectives}\label{sec:connectives}

% 首先，Isar admits 两种经常被混合使用的 interchangeable proof ideioms: structured proofs and tactic-based apply sequence

As illustrated in \cref{fig:messy-isar}, Isar uses a variety of connectives such as \isar{then}, \isar{moreover}, \isar{finally}\!, and \isar{hence} to chain proof steps within a proof block.
Semantically, they pass previously derived lemmas or conclusions as arguments to subsequent proof methods. Hence, all such connectives can be uniformly expressed as \isar{using} \texttt{<lemmas>} \isar{by} \texttt{<method>}.
For instance, \cref{fig:no-connective} shows the example from \cref{fig:messy-isar} rewritten in this normalized form.

\subsubsection{Redundant Syntax}\label{sec:redundent-syntax}
As mentioned in \cref{sec:motivation}, Isar incorporates substantial syntactic redundancy --- numerous ways to perform the same logical operation. 
Concrete examples include that Isar supports both forward, declarative proof construction (\cref{sec:isar-proof-block}) and backward, tactic-based reasoning.
Although the \isar{proof}–\isar{qed} structure already supports opening proof contexts in the declarative style, Isar supplies the \isar{subgoal} command to provide essentially equivalent support in the tactic style (\cref{fig:subgoal-to-proof-qed}).
%As illustrated in \cref{fig:subgoal-to-proof-qed}, all \isar{subgoal} statements can be transformed into \isar{proof}-\isar{qed}.
Moreover, even within the declarative style, Isar offers redundant mechanisms for opening proof contexts. An instance is the block syntax (\cref{fig:bracket-to-proof-qed}), whose effect is equivalent to a standard \isar{proof}–\isar{qed} block.
In \cref{sec:normalization}, we devote significant effort to normalizing these variants.

\subsection{The Translation Process}\label{sec:translation-detail}

The translation process includes \xqym{26} passes \xqym{(c.f. Appendix \ref{appendix:all_passes})}, each of which represents one step from an intermediate language to another closer to Minilang.
These passes fall into four stages:
\begin{enumerate}
\item Parsing the given Isar proof script into an Abstract Syntax Tree (AST).
\item Elaborating Isar’s syntactic sugars \& normalizing redundant proof idioms. This phase corresponds to the first two translation strategies mentioned in \cref{sec:introduction} (page \hyperlink{page.2}{2}).
\item Translating into Minilang (mapping the Isar AST into Minilang AST).
\item Refinement: eliminating tactics successively with Sledgehammer*. This corresponds to the last strategy in \cref{sec:introduction} (page \hyperlink{page.2}{2}).
\end{enumerate}
We discuss each stage in the subsections that follow.

\subsubsection{Parsing}

Isar is implemented in Isabelle/ML, and so is our translator. This allows us to reuse Isar’s parser to extract ASTs, ensuring correct handling of Isar’s various syntactic corner cases.
Concretely, we reuse Isar’s lexer in full and modify the final stage of the parser to emit our custom AST.
This custom AST comprises 42 node kinds, covering a broad range of Isar constructs.

\subsubsection{Elaboration \& Normalization}\label{sec:normalization}

The stage (i) elaborates syntax sugars, making implicit information explicit, and (ii) normalizes equivalent or near-equivalent ways of proof formalization into a uniform representation. This reduces the AST from 42 node kinds to 15, simplifying the subsequent translation to Minilang.

Many similar ways of proof formalization are not strictly equivalent: they coincide in most situations but diverge in certain corner cases, exhibiting subtle semantic differences. Our normalization attempts to account for these cases, but cannot cover them exhaustively. Consequently, normalization may change the semantics of a proof script. We therefore re-run Isabelle’s proof checker to ensure that the modified proofs remain valid, and discard any that fail. This normalization step contributes a major source of our translation failures.

Major steps in the elaboration \& normalization process are presented as follows, starting with localized transformations:
\begin{itemize}
\item Normalize all connectives into \isar{using} with the lemmas' explicit names.
%all connectives can be normalized into \isar{using}.
%In this pass,
%our translator normalizes connective-based chaining into an explicit \texttt{using} form with named lemmas.
As mentioned in \cref{sec:connectives}, this is implemented by rewriting rules such as
\begin{gather*}
\text{\isar{have A} ... \isar{then have B apply m} $\equiv$ \isar{have A} ... \isar{have B using A apply m}}\\[-.25em]
\text{\isar{with A have B apply m} $\equiv$ \isar{have B using this A apply m}}
\end{gather*}
\item Assign anonymous lemmas with generated names. This prepares the next pass.
\item Resolve pronouns \texttt{this}, \texttt{that}, \texttt{prems}, and \texttt{assms}.
Depending on context, these pronouns refer to the recent lemmas or assumptions. Our translator replays the Isar proof to obtain each pronoun’s binding context, uses Isabelle/ML internals to resolve the reference, and replaces it with the generated name from the previous pass, making all references explicit.
\item Unfolding the macro variables \texttt{?thesis} and \texttt{?case} that refer to proof goals in local contexts.
We use the same approach as above, invoking Isabelle's internal interfaces to resolve these references and replace them with explicit terms.

%but Minilang has no such notion.
%This pass replays the Isar script to obtain the contexts of the macro variables; then use Isabelle/ML’s internal interfaces to dereference them.
%Isar uses a variety of connectives to chain proof steps; semantically, they pass previously derived lemmas as arguments to subsequent proof methods. Hence, all such connectives can be uniformly expressed as \isar{using} \texttt{<lemmas>} \isar{apply} \texttt{<method>}.
\item Isar allows lemmas to be referenced either by name or by their full expressions. This pass replaces expression-based references with name-based ones.
\item Add type annotations to variables and numbers. Although Isabelle’s type inference works well in most cases, it sometimes fails to infer types for certain variables or numbers, causing proofs to fail. In this pass, the translator explicitly annotates all variables and numerals with their types and retains these annotations in the final training corpus, so the model learns to emit the same type markings at generation time and thus avoids such failures.
%We implement this by replaying the proof to obtain the relevant context and using Isabelle/ML internals to locate the corresponding lemma names.
\end{itemize}

Next are transformations on proof structure:
\begin{itemize}
%\newsavebox{\bracketBOX}
%\newsavebox{\bracketBOXB}

\begin{lrbox}{\bracketBOX}
\begin{minipage}{2.6cm}\centering
\begin{lstlisting}[language=isar, style=XXX, basicstyle=\ttfamily\small]
{
  fix &$x$&
  assume &$A$&
  have &$B_1$& ... &$B_n$&
  have &$C$& by &$u$&
}
\end{lstlisting}
\end{minipage}
\end{lrbox}

\begin{lrbox}{\bracketBOXB}
\begin{minipage}{2.8cm}\centering
\begin{lstlisting}[language=isar, style=XXX, basicstyle=\ttfamily\small]
have &$C$& if &$A$& for &$x$&
proof -
  have &$B_1$& ... &$B_n$&
  show &$C$& by &$u$&
qed
\end{lstlisting}
\end{minipage}
\end{lrbox}

\begin{lrbox}{\subgoalBOX}
\begin{minipage}{2.8cm}\centering
\begin{lstlisting}[language=isar, style=XXX, basicstyle=\ttfamily\small]
&\it current goal: $\bar{H} \longrightarrow G$&
subgoal
  premises &\normalfont\textit{name}&
  for &\normalfont\textit{variables}&
by &\normalfont\textit{tactics}&
\end{lstlisting}
\end{minipage}
\end{lrbox}

\begin{lrbox}{\subgoalBOXresidue}
\begin{minipage}{4cm}\centering
\begin{lstlisting}[language=isar, style=XXX, basicstyle=\ttfamily\small]
proof -
  fix &\normalfont\textit{variables}&
  assume &\normalfont\textit{name}&: &$\bar{H}$&
  show &$G$& by &\normalfont\textit{tactics}&
qed
\end{lstlisting}
\end{minipage}
\end{lrbox}

\begin{figure}
\centering
\begin{minipage}{.47\textwidth}
\[ \usebox{\subgoalBOX} \longrightarrow\ \usebox{\subgoalBOXresidue} \]
\captionof{figure}{Transforming \isar{subgoal} into \isar{proof}-\isar{qed}.}
\label{fig:subgoal-to-proof-qed}
\end{minipage}%
\hspace*{.025\textwidth}%
\rule[-0.7cm]{0.5pt}{2cm}%
\hspace*{.023\textwidth}%
\begin{minipage}{.47\textwidth}
\[\usebox{\bracketBOX}\ \ \equiv\ \ \usebox{\bracketBOXB}\]
\vspace*{-1.25em}
\captionof{figure}{Rewriting bracket block into \isar{proof}-\isar{qed}.}
\label{fig:bracket-to-proof-qed}
\end{minipage}%
%\caption{Two redundant ways of introducing lemmas in Isar}
%\label{fig:enter-label}
\Description{}
\end{figure}
%\newsavebox{\reorgboxA}
%\newsavebox{\reorgboxB}
%\newsavebox{\reorgboxC}

\begin{lrbox}{\reorgboxA}
\begin{minipage}{2.7cm}\centering
\begin{lstlisting}[language=isar,
style=XXX,
basicstyle=\ttfamily\small
]
proof
  have &$A_1$& ... &$A_n$&
  show &$G_3$& and &$G_1$&
    by &$\mathit{tac}_A$&
next
  show &$G_2$& by &$\mathit{tac}_B$&
qed
\end{lstlisting}
\end{minipage}
\end{lrbox}

\begin{lrbox}{\reorgboxB}
\begin{minipage}{2.7cm}\centering
\begin{lstlisting}[language=isar,
style=XXX,
basicstyle=\ttfamily\small
]
proof
  have &$A_1$& ... &$A_n$&
  show &$G_3$&
    by &$\mathit{tac}_A/G_3$&
  show &$G_1$&
    by &$\mathit{tac}_A/G_1$&
next
  show &$G_2$& by &$\mathit{tac}_B$&
qed
\end{lstlisting}
\end{minipage}
\end{lrbox}

\begin{lrbox}{\reorgboxC}
\begin{minipage}{3.1cm}\centering
\begin{lstlisting}[language=isar,
style=XXX,
basicstyle=\ttfamily\small
]
proof
  have &$A_1$& ... &$A_n$&
  show &$G_3$& by &$\mathit{tac}_A/G_3$&
next
  have &$A_1$& ... &$A_n$&
  show &$G_1$& by &$\mathit{tac}_A/G_1$&
next
  show &$G_2$& by &$\mathit{tac}_B$&
qed
\end{lstlisting}
\end{minipage}
\end{lrbox}

\begin{lrbox}{\reorgboxD}
\begin{minipage}{3.1cm}\centering
\begin{lstlisting}[language=isar,
style=XXX,
basicstyle=\ttfamily\small
]
proof
  have &$A_1$& ... &$A_n$&
  show &$G_1$& by &$\mathit{tac}_A/G_1$&
next
  show &$G_2$& by &$\mathit{tac}_B$&
next
  have &$A_1$& ... &$A_n$&
  show &$G_3$& by &$\mathit{tac}_A/G_3$&
qed
\end{lstlisting}
\end{minipage}
\end{lrbox}

\begin{figure}[t]
    \centering
\begin{tikzpicture}
\node[inner sep=2pt] (codeA) {\usebox{\reorgboxA}};
\node[inner sep=2pt, right=3mm of codeA] (codeB) {\usebox{\reorgboxB}};
\node[inner sep=2pt, right=3mm of codeB] (codeC) {\usebox{\reorgboxC}};
\node[inner sep=2pt, right=3mm of codeC] (codeD) {\usebox{\reorgboxD}};

\coordinate (codeA-bottom-3of4) at ($(codeA.south west)!0.75!(codeA.south east)$);
\coordinate (codeB-bottom-1of4) at ($(codeB.south west)!0.25!(codeB.south east)$);
\coordinate (codeB-bottom-3of4) at ($(codeB.south west)!0.75!(codeB.south east)$);
\coordinate (codeC-bottom-1of4) at ($(codeC.south west)!0.25!(codeC.south east)$);
\coordinate (codeC-bottom-3of4) at ($(codeC.south west)!0.75!(codeC.south east)$);
\coordinate (codeD-bottom-1of4) at ($(codeD.south west)!0.25!(codeD.south east)$);

\draw[->,thick]
  ($(codeA.south)$)
  -- ++(0,-8mm)
  -| (codeB-bottom-1of4)
  node[pos=.25, below] {\small split multi-goal \isar{show}};
  
\draw[->,thick]
  ($(codeB-bottom-3of4)$)
  -- ++(0,-3.6mm)
  -| (codeC-bottom-1of4)
  node[pos=.25, below] {\parbox{4cm}{\small\centering duplicate blocks\\containing multiple \isar{show}}};

\draw[->,thick]
  ($(codeC-bottom-3of4)$)
  -- ++(0,-3.6mm)
  -| (codeD-bottom-1of4)
  node[pos=.25, below] {\small reordering};

%\draw[->, thick] ($(codeB.east)$) -- ($(codeC.west)$) node[midway, above] {\parbox{1.5cm}{\small\centering duplicate\\goal \isar{show}}};
\end{tikzpicture}
\caption{Restructuring Isar's \isar{proof}-\isar{qed} blocks to align with Minilang's one-block-per-subgoal organization. Notation $\mathit{t}/G_i$ denotes the tactic $t$'s segment for goal $G_i$}
    \label{fig:reorganization}
    \Description{}
\end{figure}
\item Reorganize Isar's \isar{proof}-\isar{qed} structure to align with Minilang's one-block-per-subgoal organization. This reorganization process is illustrated in \cref{fig:reorganization}. As noted in \cref{sec:isar-proof-block}, a single Isar proof-qed block may contain several \isar{next}-delimited sub-blocks; each sub-block may include multiple \isar{show} statements; and each \isar{show} may address multiple goals.
To reorganize this nested structure, the translator first splits each multi-goal \isar{show} into single-goal \isar{show} statements.
The split also applies to the statement's associated tactic sequence: our translator tracks the effect of each tactic application to determine which goal it addresses, and duplicates any tactic that affects multiple goals.
Next, any proof block that contains multiple \isar{show} statements is duplicated (as illustrated in \cref{fig:reorganization}) so that each resulting block contains exactly one \isar{show}.
In addition, a dependency analysis is performed to remove unused lemmas from the duplicated blocks.
Finally, the blocks are reordered to follow the sequence of the proof goals.

%上面的改得很好了

%Then this step duplicates proof blocks containing multiple \isar{show} statement so that each block contains exactly one \isar{show}.
%Finally, the step reorders the blocks so that 它们与 proof goals 的顺寻一致。

%Isar's \isar{proof}-\isar{qed} structure is flexible, which 

%\isar{proof}-\isar{qed} blocks can contain multiple \isar{next}-separated sub-blocks, each of which may contain multiple \isar{show} statements, and each statement may target multiple goals. These elements can appear in arbitrary orders. We normalize the structure so that each \isar{proof} block targets exactly one goal, and all blocks are ordered consistently with their corresponding goal occurrences.

%\newsavebox{\subgoalBOX}
%\newsavebox{\subgoalBOXresidue}

\item Normalize each \isar{subgoal} statement into \isar{proof}-\isar{qed} block using the transformation in \cref{fig:subgoal-to-proof-qed}.
\item Rewrite each bracket structure into a \isar{proof}-\isar{qed} block (\cref{fig:bracket-to-proof-qed}).
\item Normalize tactic \texttt{goal\_cases} into the standard \isar{proof} - \isar{case} - \isar{qed} structure. %The detail is left in our code.
%As mentioned in \cref{sec:redundent-syntax}, \isar{subgoal} performs functions similar to \isar{proof}-\isar{qed}.
%This pass eliminates \isar{subgoal} statements by the following transformation:
%\begin{tikzpicture}
%\node[inner sep=2pt] (codeA) {\usebox{\reorgboxA}};
%\end{tikzpicture}

%Isar admits interchangeable structured proofs and tactic-based apply sequences. In the apply style, a \texttt{subgoal} statement opens a local context, fixing the universally quantified variables and bringing the hypotheses from the goal proposition into the context.
%In this step, the translator replaces each \texttt{subgoal} statement with the corresponding structured \isar{proof}-\isar{qed} block. Since \texttt{subgoal} fixes all hypotheses and universally quantified variables into the local context under a designated name.
%The bindings that subgoal would introduce are made explicit by \isar{fix} for variables and \isar{assume} for hypotheses.

%\item Definition unfolding can be invoked by either tactic \texttt{unfold\_tac} or command \isar{unfolding}. This step normalizes them to \texttt{unfold\_tac}.

\item In Isar, variable declarations by \isar{fix} and assumption declarations by \isar{assume} may appear anywhere within a \isar{proof}-\isar{qed} block, interleaved with other proof operations. This pass normalizes that freedom by moving all such declarations to the beginning of their containing block.
Additionally, not all variables and assumptions in Isar are declared explicitly via \isar{fix}/\isar{assume}; some are introduced implicitly within \isar{show} statements. This pass lifts such implicit introductions to explicit declarations at the block header. Consequently, each \isar{proof}–\isar{qed} block is normalized into a form amenable to direct translation to Minilang’s \minilang{INTRO} command, via the mapping in \cref{fig:translating-rules} as discussed later in \cref{step:translate}.

%\item Variables and hypotheses are \emph{selectively} fixed into the context in arbitrary order in Isar. This step normalizes this by requiring \emph{all} variables and hypotheses to be declared at the proof body beginning in their occurrence order.

%\item Tactics \texttt{induct\_tac}, \texttt{induct} and \texttt{case\_tac} are subtle variants of \texttt{induction} and \texttt{case}. This step normalizes the former into the latter.

\end{itemize}

Finally, we consider elaboration and normalization of tactic sequences:
\begin{itemize}
\item Tactic application in Isar may appear in different syntax (\isar{proof}, \isar{qed}, \isar{apply}, and \isar{by}).
This pass rewrites these occurrences so that every tactic application is expressed uniformly as \isar{apply}, allowing subsequent passes to consider \isar{apply} only.
This rewrite is implemented by rules like,
$\text{\isar{proof} (\textit{tactic}) $\equiv$ \isar{apply} (\textit{tactic}) \isar{proof -}}
\ \ \text{, and}\ 
\text{\isar{by} (\textit{tactic})} \equiv \text{\isar{apply} (\textit{tactic}) \isar{done}}.$

\item Isar tactics can be combined by combinators (\small\texttt{,\,+\,?\,|\,[$n$]}\normalsize). This step eliminates combinators and normalizes composite tactics into sequences of atomic tactics.
This is implemented by tracing the execution of composite tactics and expanding each composite into the exact sequence of atomic invocations that actually run.

\item Isar tactics may operate on multiple goals (e.g., \texttt{auto} affects all open goals). By contrast, Minilang requires operations for different proof goals to be delimited by \minilang{NEXT}; a single operation is disallowed to range over multiple goals.
As a result, the translation must identify, for each goal, the sequence of tactics that act on it.
This is implemented by tracing every tactic invocation and recording its target goals; if an invocation spans several goals, we duplicate it once per goal and confine each copy to that goal.

The resulting per-goal sequences are represented as an AST node $\textbf{ForGoals}(\mathit{seq}_1,\cdots,\mathit{seq}_n)$, where each $\mathit{seq}_i$ can only be a sequence of \isar{apply} and/or \textbf{ForGoals}.
This ForGoals AST node is subsequently translated into Minilang's \minilang{NEXT} in \cref{step:translate} by the rule in \cref{fig:translating-rules}.

%Isar tactics can act on multiple goals (e.g., \texttt{auto} affects all goals), which obscures the correspondence between tactics and individual proof goals at the syntactic level.
%This pass makes the correspondence syntactically explicit by identifying, for each goal, the exact sequence of tactics that act on it.
\end{itemize}

\subsubsection{Translating into Minilang}\label{step:translate}

% \newsavebox{\transBoxA}

\begin{lrbox}{\transBoxA}
\begin{minipage}{2.1cm}\centering
\begin{lstlisting}[language=isar,
style=XXX,
basicstyle=\ttfamily\small
]
fix &$\mathit{vars}$&
assume &$\mathit{assms}$&
&$\mathit{operations}$&
show &$\mathit{goal}$&
&\it tactic sequence&
\end{lstlisting}
\end{minipage}
\end{lrbox}

\begin{lrbox}{\transBoxB}
\begin{minipage}{2.2cm}\centering
\begin{lstlisting}[language=minilang,
style=XXX,
basicstyle=\ttfamily\small
]
INTRO
&$\translate{\mathit{operations}}$&
&$\translate{\mathit{tactic\;sequence}}$&
\end{lstlisting}
\end{minipage}
\end{lrbox}

\begin{lrbox}{\transBoxC}
\begin{minipage}{1.9cm}\centering
\begin{lstlisting}[language=isar,
style=XXX,
basicstyle=\ttfamily\small
]
proof -
  &$\mathit{operations}_1$&
next
  ...
next
  &$\mathit{operations}_n$&
qed
\end{lstlisting}
\end{minipage}
\end{lrbox}

\begin{lrbox}{\transBoxD}
\begin{minipage}{1.75cm}\centering
\begin{lstlisting}[language=minilang,
style=XXX,
basicstyle=\ttfamily\small
]
&$\translate{\mathit{operations}_1}$&
NEXT
&$\cdots$&
NEXT
&$\translate{\mathit{operations}_n}$&
\end{lstlisting}
\end{minipage}
\end{lrbox}

\begin{lrbox}{\transBoxE}
\begin{minipage}{1cm}\centering
\begin{lstlisting}[language=minilang,
style=XXX,
basicstyle=\ttfamily\small
]
&$\translate{\mathit{seq_1}}$&
NEXT
&$\translate{\cdots}$&
NEXT
&$\translate{\mathit{seq_n}}$&
\end{lstlisting}
\end{minipage}
\end{lrbox}

\begin{figure}[t] \centering\small
\begin{minipage}{.54\textwidth}
\begin{gather*}
\translate{\usebox{\transBoxA}} \ \triangleq\  \left(\ \usebox{\transBoxB}\ \right)
\\[.2em]
\translate{\text{\isar{apply} ($\mathit{tactic}$)}} \triangleq \text{\minilang{APPLY} ($\mathit{tactic}$)}
\qquad
\translate{\text{\isar{done}}} \triangleq \text{\minilang{END}}
\end{gather*}
\end{minipage}\hspace*{.01\textwidth}%
\begin{minipage}{.45\textwidth}
\[
\translate{\usebox{\transBoxC}} \ \triangleq\  \left(\ \usebox{\transBoxD}\ \right)
\]
\end{minipage}
\begin{gather*}
%tactic sequence
\translate{\text{\isar{have} $\mathit{propositions}$}} \triangleq \text{\minilang{HAVE} $\mathit{propositions}$}
\qquad\qquad
\translate{\text{\isar{consider} $\mathit{cases}$}} \triangleq \text{\minilang{CONSIDER} $\mathit{cases}$}
\\[.2em]
\translate{\text{\isar{obtain} $\mathit{vars}$ \isar{where} $\mathit{conditions}$}} \triangleq \text{\minilang{CONSIDER} $\exists\mathit{vars}.\;\mathit{conditions}$}
\\[.2em]
\translate{\textbf{ForGoals}(\mathit{seq}_1,\mathit{seq}_2,\cdots,\mathit{seq}_n)} \triangleq \text{$\translate{\mathit{seq}_1}$ \minilang{NEXT} $\translate{\mathit{seq}_2}$ \minilang{NEXT} $\cdots$ \minilang{NEXT} $\translate{\mathit{seq_n}}$}
\end{gather*}\normalsize
    \caption{Selected translation rules from Isar to Minilang. $\translate{\cdot} : \mathrm{Isar} \rightarrow \mathrm{Minilang}$ is the translation mapping.}
    \Description{}
    \label{fig:translating-rules}
\end{figure}

By this stage, most redundant Isar constructs have been eliminated, and the proof scripts have been normalized.
From their normalized forms, a small set of mapping rules suffices to translate the scripts into Minilang.
A subset of these rules is shown in \cref{fig:translating-rules}.

Note that tactic applications have not yet been eliminated at this stage. To express them in Minilang, we extend Minilang with an \minilang{APPLY} command to denote a tactic application. Its semantics is defined as follows, which mirrors Isar’s \isar{apply} but acts only on the first subgoal.
\small
\begin{center}
\begin{tikzpicture}
  \node (LEFT) {%
    \begin{forest} for tree={align=center, l sep=0, s sep=0}
      [$\mathcal{R}$
        [$\CCTX{(\Theta, \Gamma)}\vdash \CGOAL{G_0}$]
        [$\mathcal{S}$]
      ]
    \end{forest}
  };

  \node (RIGHT) [right=8mm of LEFT] {%
    \begin{forest} for tree={align=center, l sep=0.15cm, s sep=2mm}
      [$\mathcal{R}$
        [$\CCTX{(\Theta,\Gamma)}$
          [$\CCTX{(\varnothing,\varnothing)}\vdash\NOTE{G_1}$]
          [$\CCTX{\cdots}$]
          [$\CCTX{(\varnothing,\varnothing)}\vdash\NOTE{G_n}$]
        ]
        [$\mathcal{S}$]
      ]
    \end{forest}
  };
  
  \draw[->, thick] ($(LEFT.east)+(-1mm,0)$) -- ($(RIGHT.west)+(15mm,0)$) node[midway, above] {\minilang{APPLY} $\mathit{tactic}$};
  
  \node[right=1mm of RIGHT] {\parbox{45mm}{if applying $\mathit{tactic}$ to $\mathit{G_0}$ yields\\subgoals $G_1,\cdots,G_n$}};
\end{tikzpicture}
\end{center}\normalsize

%Regarding Isar's calculational reasoning (by keywords \isar{also} and \isar{finally}), we extend Minilang with a transparent automatic mechanism.
%Isar's calculation is about extending (order-theoretical) chains, e.g., using $A_n \leq A_{n+1}$ to extend a chain $A_1 \leq \cdots \leq A_n$ to derive $A_1 \leq A_{n+1}$.
%Minilang maintains a series of chains internally. Once a proposition $P$ is proven, Minilang checks if $P$ can extend any existing chain at its tail. If so, all feasible chains are extended; otherwise, $P$ initiates a new chain of length 1 in the internal series.
%Once a chain $\{A_i\}_n$ reaches a length of 2 or more, indicating a derivation has occurred, the calculation result $A_1 \leq A_n$ is added into the context or updated if it is already there.
%
%This allows users to reference the calculated result without explicit uses of \texttt{also} and \texttt{finally}.
%Though not complete, this approach can replace many uses of \texttt{also} and \texttt{finally}, reducing the concepts involved in MiniLang.

Additionally, to translate as much Isar as possible into Minilang, we extend Minilang with two auxiliary commands: one configures Isabelle’s attribute system (e.g., to register local simplification rules), and the other activates Isabelle’s bundle and module mechanisms.
These commands account for ${\sim}1.01\%$ of the translated code, and we defer their details to our supplementary materials.

%By this stage, most redundant Isar constructs have been eliminated, and the proof script has been normalized. From this normal form, a small collection of mapping rules suffices to translate the script into Minilang; a representative subset appears in Fig.~12.

%most redundant Isar statements have already been eliminated. The remaining normalized statements are translated into their corresponding Minilang equivalents: \isar{obtain} and \isar{consider} $\mapsto$ \minilang{CONSIDER}, \isar{apply} $\mapsto$ \minilang{APPLY}, \isar{done} $\mapsto$ \minilang{END} or \minilang{NEXT}, \isar{have} $\mapsto$ \minilang{HAVE}, \isar{proof} $\mapsto$ \minilang{INTRO} ... \label{step:trans}

\subsubsection{Refinement}\label{step:refinement}
%Tactic applications still remain after step~\ref{step:translate}.
%To obtain declarative proofs as pure as we can, we eliminate tactics by repeating the following substitution until \minilang{END} cannot prove the goal alone using the $\mathit{ATP}$.
Tactic applications still remain after the translation step in \ref{step:translate}.
Since Minilang follows a purely declarative paradigm (\cref{sec:motivation}) that discourages the use of tactics, the final stage of the translation process aims to replace these remaining tactics with Sledgehammer* wherever possible.

By this stage, Minilang scripts express all tactic applications as sequences 
of \minilang{APPLY} commands that terminate with \minilang{END} or \minilang{NEXT}.
Attempting to eliminate these tactics, we repeatedly apply the following substitutions, which progressively absorbs each \minilang{APPLY} command immediately preceding an \minilang{END} or \minilang{NEXT} into that terminal statement.
\[
\begin{array}{lcl}
\text{\small\minilang{APPLY} $\mathit{tactic}$ \minilang{END WITH} $w$ \minilang{WITHOUT} $\mathit{wo}$}
&\mapsto & \text{\small\minilang{END WITH} $w$ $\mathit{lem}^+$ \minilang{WITHOUT} $wo$ $\mathit{lem}^-$}
\label{eq:sub-tactics}\\[-.1em]
\text{\small\minilang{APPLY} $\mathit{tactic}$ \minilang{NEXT} $w$ \minilang{WITHOUT} $\mathit{wo}$}
&\mapsto & \text{\small\minilang{NEXT WITH} $w$ $\mathit{lem}^+$ \minilang{WITHOUT} $wo$ $\mathit{lem}^-$}\\
&&\hspace*{-9.7em}\textit{if, Sledgehammer* can still prove the goal after removing the tactic}
\end{array}
\]

Recall from §\ref{sec:sledgehammer} that a key innovation of this work is that we extract both relevant lemmas and lemmas to avoid as annotations into the training corpus, enabling proof generation models to learn how to suggest these hints to assist Sledgehammer*'s premise selection.
In the substitution rule above, $\mathit{lem}^+$ and $\mathit{lem}^-$ represent respectively the relevant and avoided lemmas extracted from the arguments of $\mathit{tactic}$. An example is \texttt{auto add:} $\mathit{lem}^+$ \texttt{del:} $\mathit{lem}^-$.
These lemmas are merged into the \minilang{WITH} and \minilang{WITHOUT} clauses as hints, where $w$ and $\mathit{wo}$ represent the lemma hints already present before the substitution.

The extraction of the $\mathit{lem}^+$ and $\mathit{lem}^-$ is performed by a heuristic we developed.
It recognizes the argument syntax of 44 common Isabelle tactics to identify relevant/avoided lemmas, and defaults to treating all lemma arguments as relevant for unrecognized tactics.

The substitutions continue until either no \minilang{APPLY} remains or further substitution would cause \minilang{END} or \minilang{NEXT} to fail to prove the goal using the Sledgehammer*.
In the latter case, we are forced to retain the remaining \minilang{APPLY} commands in the training corpus, which represents a compromise to the purely declarative ideal advocated in §\ref{sec:minilang}.
Fortunately, with the aid of relevant lemma annotations, ${>}95\%$ of tactics are successfully eliminated, leaving APPLY commands accounting for only 6.7\%  of all translated Minilang commands.

As an exception, when tactics correspond to common proof operations that Minilang directly supports, the tactics are substituted into the corresponding commands instead of \minilang{END} or \minilang{NEXT},
%
%Additionally, some common tactics have specialized substitutions mapping them into specific Minilang statements, e.g.,
%
\begin{align*}
\text{\small\minilang{APPLY} (\texttt{auto} $\mathit{args}$)} &\mapsto \text{\small\minilang{SIMPLIFY} $\mathit{args}$}&
\text{\small\minilang{APPLY} (\texttt{simp} $\mathit{args}$)} &\mapsto \text{\small\minilang{SIMPLIFY} $\mathit{args}$}\\[-2pt]
\text{\small\minilang{APPLY} (\texttt{cases} $\mathit{args}$)} &\mapsto \text{\small\minilang{CASE_SPLIT} $\mathit{args}$}&
\text{\small\minilang{APPLY} (\texttt{induct} $\mathit{args}$)} &\mapsto \text{\small\minilang{INDUCT} $\mathit{args}$}
\end{align*}

\section{Evaluation}\label{sec:evaluation}
%\conrad{translation results should be reported as part of eval section, not above}
%\DS{The research questions from the introduction should be here, and the results should clearly show (as much as possible) the answers to the questions}

%In this section, we finally evaluate the performance of MiniLang applied in NTP. To conduct such an evaluation, we first build a typical NTP system over MiniLang.

The translation method in \cref{sec:translation} provides a way to construct Minilang corpora for training language models. This allows us to evaluate Minilang's effectiveness for NTP through fine-tuning LLMs. In this section, we conduct an ablation study by fine-tuning 2 pretrained LLMs under 3 ATP configurations (Sledgehammer*, original Sledgehammer, and no ATP) for both Minilang and Isar, yielding 12 models in total to compare their pass rates on the same benchmark (PISA~\cite{PISA}).

%In this section, we evaluate Minilang's effectiveness for NTP through fine-tuning LLMs.
% To this end, we first construct Minilang's training corpora by translating AFP using the pipeline established in \cref{sec:translation}.
% \cref{sec:translation-result} analyzes this translation result.
% 
% §5.1 analyzes the translation results and corpus characteristics. §5.2 presents our experimental design: we conduct an ablation study comparing MiniLang against Isar across six different configurations, fine-tuning twelve language models in total. Finally, §5.3 analyzes the experimental results to answer our research questions about MiniLang's effectiveness.
% 
% Having obtained a substantial corpus through translation, we evaluate the impact of proof language redesign on NTP performance by building and comparing NTP systems over both Isar and MiniLang.

Based on the evaluation results, we answer the following research questions:
\begin{questions}
\item Can we effectively improve declarative NTP by redesigning the underlying proof language?
%If so, how effective can it be?
%If so, how effective is redesigning the underlying proof language?
\item Is improvement confined to syntax error reduction, or are reasoning errors also reduced?
%\item What can be an effective principle to guide the design of a proof language for declarative NTP?
\item For a redesigned proof language, how can we obtain effective training corpora for it?
\end{questions}

In what follows, \cref{sec:translation-result} analyzes the translation results and the corpus features; \cref{sec:NTP-setup} details the ablation study design and fine-tuning configurations; \cref{sec:results} analyzes the experiment results.

\subsection{Analysis of the Translation Results}\label{sec:translation-result}

As discussed in Section 4.2.2, Isar's language features encompass numerous corner cases that our translator cannot exhaustively cover. Additionally, Isar is extensible, and many works in AFP define custom commands that fall outside our translator's scope.
These factors limit our translation success rate to ${\sim}85.28\%$. Nevertheless, the resulting 290K proofs are still a substantial corpus sufficient for supervised fine-tuning.
%AFP (version 2025-02-12) includes ${\sim}324k$ proofs, and the translation yields ${\sim}276k$ MiniLang proofs, still a substantial corpus sufficient for supervised fine-tuning.

%TODO: How many CPU hours are taken???

%To examine the degree of simplification achieved by Minilang over Isar, we analyze translation outcomes across Isar proofs of different sizes (measured by command count). Figure~\ref{fig:simplification} shows both the translation success rate and the ratio of command counts between translated Minilang proofs and their original Isar counterparts.

\begin{wrapfigure}{r}{0.48\textwidth}
  \centering
  \includegraphics[width=\linewidth]{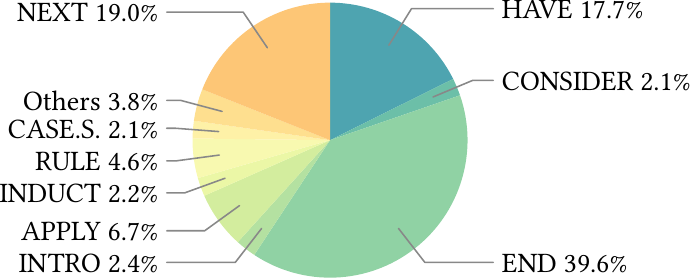}
  \caption{Distribution of Minilang commands in the translation result.}
  \Description{}
  \label{fig:translation-distr}
\end{wrapfigure}

%The failed proofs are discarded, and we finally obtain 85.28\% proofs of the entire AFP.
To examine the quality of the translation, we analyze the occurrence frequency of every command in the translation result.
The result is shown in \cref{fig:translation-distr}, where \minilang{END} and \minilang{NEXT} are the most frequent commands. This is because (1) every goal and subgoal must be closed by END or NEXT,
and (2) 42.65\% of Isar proofs can be proven directly by Sledgehammer* with a single END statement, requiring no other proof steps.

\begin{figure}[t]
    %\hspace*{-.6em}\vspace*{-0.6em}
\begin{minipage}{.49\textwidth}
\includegraphics[width=\linewidth]{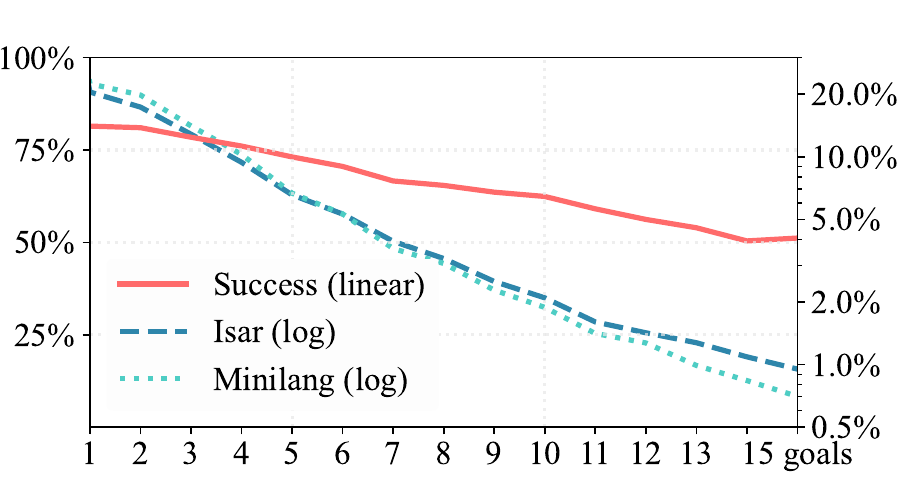}%
\captionof{figure}{Translation success rate (solid line) and distributions of 
Isar (dashed) and Minilang proofs (dotted) over the number of subgoal declarations.\\}
\label{fig:distr_have}
\end{minipage}%
\hspace*{.02\textwidth}%
\begin{minipage}{.49\textwidth}
\includegraphics[width=\linewidth]{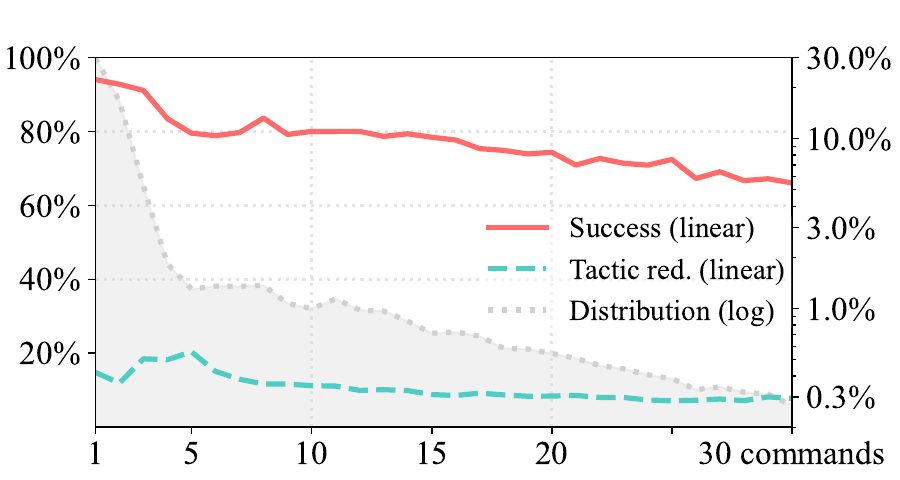}
\captionof{figure}{Translation success rate (solid line), the ratio of tactic counts after/before translation (dashed line), and the distribution of Isar proofs (blue shadow) over the number of Isar commands.}
\label{fig:translation-success}
\end{minipage}%
\Description{}
%\vspace*{.1em}
%\caption{Solid line: translation success rate on proofs containing different numbers of \isar{have} and \isar{hence} commands (cmds). Dashed and dot lines: Distributions of the number of \minilang{HAVE} and \isar{have} + \isar{hence} in the translated and the original proofs. }
%\label{fig:distr_have}
\end{figure}

Subgoal declarations (\minilang{HAVE} and \minilang{CONSIDER}) constitute the next largest portion, confirming their central role in declarative proofs.
To evaluate how well our translation preserves declarative structures, we first examine the distribution of subgoal declaration counts before and after the translation.
Specifically, Isar's count is measured by the number of \isar{have}, \isar{hence}, \isar{consider}, and \isar{obtain} commands; Minilang's is by \minilang{HAVE} and \minilang{CONSIDER}.
As shown by the dashed line (original Isar) and dotted line (translated MiniLang) in Figure~\ref{fig:distr_have}, the distributions remain nearly identical, indicating that the declarative structures are largely well preserved.

Using the number of subgoal declarations as a proxy for proof complexity, we further assess how the translation success rate varies with this complexity.
The result is illustrated by the solid line in \cref{fig:distr_have}.
For short-to-moderate declarative proofs ($\leq 5$ subgoals), the success rate exceeds 70\%. As the number of subgoal declarations increases, the success rate declines gradually; however, even for complex proofs with 15 subgoals --- which account for less than 1\% of all declarative proofs --- the success rate remains above 50\%.
This demonstrates that our translator is reasonably successful in preserving declarative proof structures, given the syntactic complexity of Isar.

As a complementary measure, we assess the translation success using the number of Isar commands as an alternative complexity metric.
As illustrated in \cref{fig:translation-success}, for moderate-sized Isar proofs  (up to 15 commands), the success rate remains above 78\%. Even for longer proofs with 30 commands, the success rate maintains 66\%. These results further confirm that our translator is capable of handling proofs of substantial length.

%However, over 93\% of proofs contain 5 or fewer such statements, where the translation success rate remains above 71.5%. This demonstrates that our translation effectively preserves the majority of declarative proofs in the corpus.
%
%As the number of \isar{have} and \isar{hence} statements increases, the success rate decreases rapidly. since more complex proofs introduce additional corner cases not covered by our translation rules. However, over 93\% of proofs contain 5 or fewer such statements, where the translation success rate remains above 71.5%. This demonstrates that our translation effectively preserves the majority of declarative proofs in the corpus.

%The next largest portion is by HAVE, which represents that HAVE is the major component in declarative proofs. Thus, in \cref{fig:distr_have}, we calculate the translation success rate on proofs containing different numbers of \isar{have} and its relative \isar{hence}, to examine the quality of the translation in preserving declarative proofs.
%The success rate decreases rapidly when the number of \isar{have} and \isar{hence} increases. \CW{reemphasise that have and hence come from input - some will survive as HAVE, some will be eliminated}
%This is because the potential corner cases not covered by the translation rules increase as the number increases.
%Nonetheless, this number is not more than 5 for most proofs ($> 93.0\%$), where the success rate still remains above 71.5\%. Thus, we believe, this translation process is acceptable in preserving most declarative proofs seen in the corpus.

At last, we evaluate the effectiveness of the refinement process (\cref{step:refinement}) in eliminating tactics.
The dashed line in Figure 16 plots the ratio of atomic tactics in the translated corpus to those in the original Isar. As it reveals, the refinement process reduces tactics by more than 80\% across proofs of all sizes.
%Additionally, the refinement process reduces \minilang{APPLY} commands from 37.2\% to 6.7\%.
This confirms the efficacy of our refinement strategy for tactic elimination.

% the portion of APPLY reduces from 37.2\% to 6.7\%.
% It means a large number of tactics that harm the purity of declarative proofs are eliminated, though some of them are still present and demand more translation rules to cover them.
% At last, the portion of APPLY reduces from 37.2\% to 6.7\%.
% It means a large number of tactics that harm the purity of declarative proofs are eliminated, though some of them are still present and demand more translation rules to cover them.
% % TODO : EXPAND TODO!!s!

\subsection{Model Training and Evaluation Setup}\label{sec:NTP-setup}

To minimize confounding factors in this evaluation, we carefully set up our experiments: (1) we use Supervised Fine-Tuning (SFT) --- a standard and basic training approach --- to train NTP models on both Isar and MiniLang; (2) we adopt whole-proof generation, where language models directly produce complete proofs without additional search frameworks or inference-time scaffolding beyond temperature sampling; (3) we follow the prompt setup from Baldur, a prior NTP work on Isabelle to eliminate variations from prompt design. The concrete setup is detailed as follows.

\subsubsection{Base Models}
To assess Minilang across multiple models, we fine-tune two 7B base models: Llemma~\cite{Llemma} and DeepSeek-Prover-Base v1.5 (DPSK-PB)~\cite{DPSK1.5}. Both models have been pre-trained on datasets from major proof assistants including Coq, Lean, and Isabelle.

\subsubsection{Data Source}\label{sec:data-source}

Our fine-tuning datasets are sourced from Isabelle AFP version 2025-02-12 \cite{AFP} and Isabelle 2024 HOL libraries \cite{isabelle-HOL-std}.
We apply several preprocessing steps to these sources: we remove unreachable code, exclude proofs of benchmark targets (such as those in PISA~\cite{PISA}), and filter out proofs that exceed the context window limits of our base LLMs.
After preprocessing, these sources yield a total of ${\sim}332$K proofs.

These preprocessed datasets are then converted into six versions, one corresponding to each experimental condition in our ablation study, as detailed in the following subsection.

\subsubsection{Abolation Setup \& Dataset Construction}

Our work introduces two primary improvements: the redesign of the proof language and the enhancement of Sledgehammer. To comprehensively evaluate their respective impacts on NTP performance, we design an ablation study covering all combinations of two language choices (Isar vs. Minilang) and three ATP configurations, Sledgehammer* (SH*), Sledgehammer (SH), and no ATP.
This constitutes $2 \times 3 = 6$ conditions, necessitating six versions of the training corpus. These conditions and their corpus are detailed as follows:
\begin{itemize}
\item \textbf{Isar + no ATP}: This evaluates the original Isar, using the corpus obtained in \cref{sec:data-source}.
\item \textbf{Isar + SH}: This evaluates \emph{Thor-style Isar} language~\cite{Thor}, which represents the best-known language for declarative NTP on Isabelle.
Thor-style Isar minimizes tactic usage by delegating proof obligations to Sledgehammer.
The corpus is constructed by exhaustively replacing tactics with Sledgehammer whenever Sledgehammer can discharge the proof obligations addressed by the original tactics.
\item \textbf{Isar + SH*}: Same as Isar + SH, but use Sledgehammer* instead. The corpus contains the relevant lemma hints extracted using the same heuristic described in \cref{step:refinement}.

\item \textbf{Minilang + no ATP}: The intermediate Minilang code obtained directly from the translation step (\cref{step:translate}) without applying the refinement step (\cref{step:refinement}). This intermediate representation preserves all tactics as \minilang{APPLY} commands and deviates from Minilang's design principle of purely declarative theorem proving.

\item \textbf{Minilang + SH}: 
This evaluates a weakened version of Minilang where the ATP backend is downgraded from Sledgehammer* to the original Sledgehammer.
Its corpus is constructed by removing the lemma hints from the Minilang + SH* corpus.

\item \textbf{Minilang + SH*}: This evaluates the full-featured MiniLang. Its corpus is constructed by translating the Isar + no ATP corpus using the pipeline established in \cref{sec:translation}.

\end{itemize}

These ablation conditions differ only in their training corpus, proof language, and ATP backend; all other experimental settings remain identical across the conditions.

%
%To enable our ablation study, these datasets are converted into six versions, one for each experimental condition listed in Table~\ref{tab:PISA}.
%For the ablation study detailed in §\ref{sec:results}, we convert these datasets into six corresponding versions, one for each experimental condition listed in Table~\ref{tab:PISA}.

%During the ablation study detailed in ~\S\ref{sec:results}, we convert this dataset into separate versions for each language configuration summarized in \cref{tab:PISA}, resulting in 6 datasets in total, so that each experiment is conducted with a model fine-tuned to that particular configuration.
%These numbers reduce to $\sim$285K goals for MiniLang due to representation change and translation loss.

\subsubsection{Benchmark}

We use PISA \cite{PISA} as our evaluation benchmark.
It originally comprises 3K goals randomly sampled from Isabelle AFP version 2022-12-06.
Due to the ongoing development of Isabelle and AFP, some goals have been moved or removed from the newer versions.
We manually updated the dataset to AFP version 2025-02-12, removing \xqym{unavailable} goals, resulting in 2,962 goals.

\subsubsection{Data Contamination}

As PISA is sampled from the open-access AFP, it faces the same data contamination risks inherent to all benchmarks derived from public datasets. Llemma explicitly reports that they removed any proofs whose names appear in PISA from their pretraining data~\cite{Llemma}. However, DPSK-PB makes no such claim in their paper~\cite{DPSK1.5}; we cannot rule out the possibility of data leakage between PISA and DPSK-PB's pretraining corpus.

Nevertheless, any such contamination would more likely inflate Isar's performance relative to Minilang, since Minilang proofs do not appear in either base model's pretraining data. Thus, if Minilang still demonstrates superior performance, data contamination concerns would minimally undermine the conclusion of the paper.

%Nevertheless, any data contamination is more likely to falsely inflate the performance of Isar compared to MiniLang, since MiniLang programs do not appear in either DPSL-PB or Llemma's training data, and we have ensured that PISA is not present in the fine-tuning MiniLang corpus.
%
%Thus, such contamination would minimally undermine the conclusion of the paper if the performance of MiniLang is significantly better than that of Isar.

\subsubsection{Prompt Setup}
\label{sec:evaluation-prompt}
%Certain declarative proofs can exceed this size as expressions of subgoals require more tokens.

Following Baldur's approach, we use a simple prompt setup with two parts: context and goal.
Context includes declarations, lemmas, and proofs immediately preceding the goal in the same file.
The goal contains the name and statement.
Given the 4K token context window for both Llemma and DPSK-PB, we reserve 2K tokens each for prompt and models' response, truncating distant content when necessary.

% We follow the same simple and imperfect prompt setup as Baldur. It contains two parts: a context and a goal.
% The context consists of declarations, lemmas, and their proofs that occur immediately before the goal in the theory file of the goal.
% The goal consists of its name and proposition.
% Both the context window of Llemma and DPSK-PB is 4K.
% We leave 2K tokens to the context and 2K to the goal.
% If the complete context exceeds the limit, we truncate the latest part (at boundaries between complete elements) near the goal.

\subsubsection{Supervised Fine-Tuning (SFT)}
We use LLaMA-Factory \cite{zheng-etal-2024-llamafactory}, a widely used LLM training framework to train Llemma and DPSK-PB with supervised fine-tuning. Both models are fine-tuned for 2 epochs with a batch size of 256. The learning rate is set to $2 \times 10^{-5}$, and linearly scaled to 0 during training. The training is run on 8 Nvidia H200 GPUs, and each takes ${\sim}12$ hours to finish. 

\begin{table}[t]
\centering
\renewcommand{\arraystretch}{1.2}
\setlength{\tabcolsep}{5pt}\small
% \begin{tabular}{c c c}
%   \bf Base Models & \bf Isar / Mini (no SH*) & \bf Isar / Mini (+ SH*)  \\
%   DPSK-PB & 40.2\%\ /\ 63.1\% & 36.1\%\ /\ 68.5\% \\
%   Llemma  & 40.2\%\ /\ 62.6\% & 37.8\%\ /\ 66.9\%
% \end{tabular}
\caption{PISA evaluation of models over MiniLang and Isar. SH = Sledgehammer, SH* = Sledgehammer*.}
\label{tab:PISA}
\begin{tabular}{l l l c c c}\toprule
\bf Base Model & \bf Language & \bf \!\!ATP & \bf pass@1 & \bf pass@4 & \bf pass@8 \\\midrule
  DPSK-PB & Minilang        & SH*   & \bf 69.1\% & \bf 76.0\% & \bf 79.2\% \\
  DPSK-PB & Minilang        & SH    &     59.9\% &     68.1\% &     72.4\% \\
  DPSK-PB & Thor-style Isar & SH*   &     63.9\% &     69.6\% &     74.3\% \\
  DPSK-PB & Thor-style Isar & SH    &     45.7\% &     54.9\% &     59.7\% \\
  DPSK-PB & Minilang        & none  &     35.5\% &     40.6\% & 44.9\% \\
  DPSK-PB & Isar            & none  &     40.2\% &     45.8\% &     50.5\% \\\midrule
  Llemma  & Minilang        & SH*   & \bf 68.0\% & \bf 74.9\% & \bf 78.9\% \\
  Llemma  & Minilang        & SH    &     58.7\% &     67.9\% &     72.2\% \\
  Llemma  & Thor-style Isar & SH*   &     63.3\% &     66.9\% &     72.1\% \\
  Llemma  & Thor-style Isar & SH    &     46.1\% &     51.9\% &     57.5\% \\
  Llemma  & Minilang        & none  &     35.2\% &     39.8\% &     44.6\% \\
  Llemma  & Isar            & none  &     38.6\% &     43.9\% &     48.6\%
  %\multicolumn{2}{l}{Baldur 8b~\cite{Baldur} + Thor} \CW{delete} & \multicolumn{2}{c}{pass@64 = 65.7\%}
%\\
% \multicolumn{2}{l}{Magnushammer + Thor~\cite{Magnushammer}} & \multicolumn{2}{c}{}
  \\\bottomrule
\end{tabular}
\end{table}

% Isar 语料库中已经有很多了，大量训练了。underfit

\subsubsection{Proof Check}

We evaluate model performance using the standard pass@$k$ metric. For each benchmark entry, we sample $k$ proof attempts from the model and verify each sample using Isabelle's proof checker. An entry is considered proved if at least one of the $k$ samples passes verification. The pass@k is then computed as the proportion of proved entries in the test set.

%We sample a model's inference $k$ times for each benchmark entry to obtain the pass@$k$. We run Isabelle to check every sample. If any of the samples passes the proof check, the entry is considered passed. The pass@k is then the portion of the passed entries in the PISA test set.

\xqym{
The evaluation is conducted on a 3-node cluster with 36 Intel i7 CPU cores, taking ${\sim}$20 hours per language for pass@8. Sledgehammer and Sledgehammer* are configured with a 30-second timeout per invocation.
}

Sledgehammer and Sledgehammer* include a self-learning system that maintains local databases of premise-goal connections.
Performance on a goal improves when the hammers have previously encountered similar problems.
For fair comparison, we reset the local database before each model evaluation.
Nonetheless, this also means our results underestimate real-world performance, where Sledgehammer* would retain contextual knowledge and perform better.

% 超时时间？？!!!

\subsubsection{Infrastructure}
The proof check process is powered by a socket-based parallel Read-Eval-Print-Loop infrastructure that we developed. It is implemented in the Isabelle/ML language~\cite{isabelle-implementation} by interfacing with Isabelle's internals.

\subsection{Results}\label{sec:results}

%\DS{No comparison with other NTP like Baldur?}
%\CW{RQs as subheadings or paragraphs, reference metrics in previous sections that support your intended conclusion, state what answer you are led to by this data}

Table~\ref{tab:PISA} presents the evaluation results on PISA. Based on these results, we answer the three research questions posed at the beginning.

%Based on the evaluation results of the SFT over Isar and MiniLang as listed in \cref{tab:PISA}, we answer the three research questions posed at the beginning.

\subsubsection{RQ 1. Effectiveness of Redesigning Proof Language}

To answer RQ 1, we compare our redesigned language, Minilang + SH*, against Thor-style Isar + SH, the best-known language for declarative NTP on Isabelle~\cite{Thor}. The results demonstrate substantial improvements of ${\sim}20$ percentage points across both base models and evaluation metrics.

Our ablation study further reveals the sources of this improvement. When Minilang and Isar use the same ATP backend (either SH or SH*), Minilang consistently outperforms Isar by at least 5 percentage points across both base models and all evaluation metrics.
This advantage grows to over 12 percentage points under SH.
Conversely, holding the language fixed and upgrading from SH to SH* yields at least 6 percentage points improvement across all models and metrics.
These results demonstrate that both Minilang's language design and the ATP improvement contribute significantly to NTP performance.
These results directly answer RQ 1:
\begin{finding}
\textbf{RA 1}: Redesigning the proof language can effectively improve declarative NTP.
\end{finding}

When SH and SH* are disabled, proofs rely completely on tactic applications, turning the scenario into tactic-based theorem proving, as opposed to declarative proving. In this case, Minilang no longer maintains its advantage and falls behind Isar by $\sim$5 percentage points. This is reasonable given that Minilang is specifically designed for declarative proofs.

In particular, Minilang restricts every tactic application to act only on the leading subgoal.
While this brings clearer proof structure and better alignment between tactics and subgoals, it also prevents models from using terminating tactics (e.g., \texttt{fastforce}, \texttt{auto}) to conclude all subgoals at once.
Instead, models must precisely understand each tactic's effect and track proof state changes --- a task that is prohibitively difficult without interactive feedback from the proof assistant, even for human users. Indeed, 37.4\% of proof failures result from premature termination: models incorrectly assume all subgoals are resolved when unproven subgoals remain.
Thus, we attribute Minilang's performance degradation to its explicit exposure of the inherent difficulties of tactic-based proving for language models.

\subsubsection{RQ 2. Sources of Performance Improvement}

\begin{figure}[t]
    \centering
    \includegraphics[width=\linewidth]{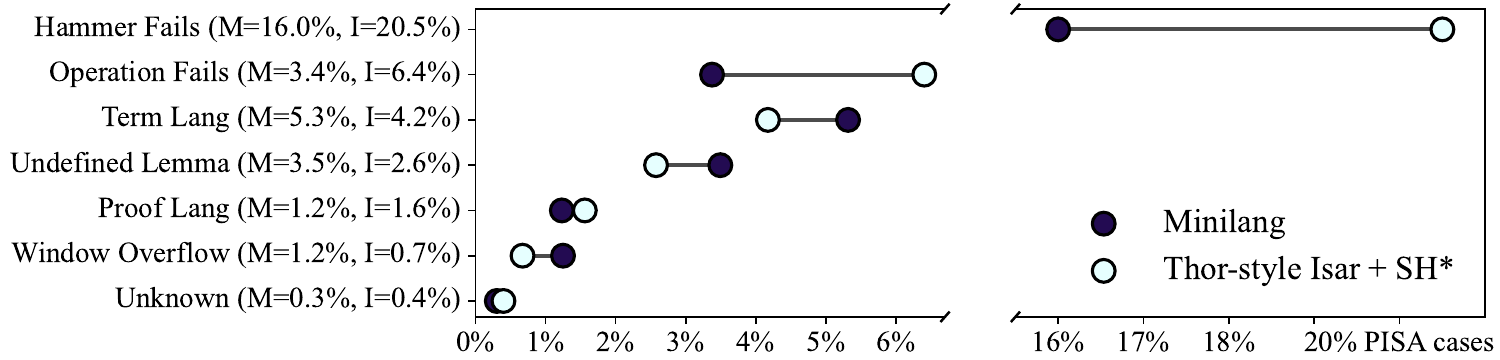}
    \caption{Comparison of failure causes between fine-tuned models over Minilang (M) and Isar+SH* (I).}
    \Description{}
    \label{fig:failures}
\end{figure}

Holding the ATP fixed at SH*, Figure~\ref{fig:failures} compares the failure causes between Minilang and Isar across both base models in the pass@1 evaluation.

Failures are categorized as follows:
(1) \emph{Hammer Fails} indicates that SH* failed to discharge a proof obligation;
(2) \emph{Operation Fails} represents failures in executing proof operations, such as tactics and other specific mechanisms like calculational reasoning;
(3) \emph{Term Lang} denotes syntax errors in term expressions;
(4) \emph{Proof Lang} captures ill-formed proof scripts;
(5) \emph{Window Overflow} arises when the allocated 2K context window is insufficient for the model to complete the proof;
%--- either because the proof is genuinely long or because the model erroneously generates repetitive output.
(6) \emph{Unknown} occurs when the evaluation infrastructure fails to capture specific error information.
%This analysis reports only the first error encountered in each failed proof.

The Proof Lang category shows that Minilang does help models generate syntactically correct proofs, but this improvement accounts for no more than 10\% of the total performance gain.

The primary improvements come from the reduced failures in proof operations and the increased hammer success rate. These gains cannot be attributed to the ATP enhancement, because both Minilang and Isar use the same SH* backend in this comparison.
Since our controlled comparison isolates language choice as the only variable, we believe these gains stem from Minilang's design helping models generate more logically valid proof steps (reducing operation failures) whose resulting proof obligations are more tractable for SH* (increasing its success rate).
%The most plausible explanation is that Minilang's design helps models generate more logically valid proof steps (reducing operation failures) and more tractable subgoals (increasing SH* success rates).
\begin{finding}
\textbf{RA 2}: The improvement brought by redesigning the proof language can extend beyond syntax errors to enhance models' capability of generating logically correct proofs
\end{finding}

% We compare the failure causes of proofs generated by Minilang and Isar + SH* in the pass@1 test.
% The result is shown in \cref{fig:failures}.
% Minilang significantly reduces syntax errors in the proofs.
% However, this accounts for no more than 10\% of the total improvement.

%More significantly, MiniLang reduces failures in proof operations and Sledgehammer* calls, meaning the generated MiniLang proofs are more likely to be correct.
%This indicates that the language design helps LLMs generate higher-quality proofs, not merely syntactically correct text.

We also observe an increase in term language errors for Minilang, despite no modifications to this aspect of the language. This likely reflects the sequential nature of our error analysis, which reports only the first error encountered --- as Minilang resolves certain types of errors, other unrelated issues may become more visible in the error categorization.

% \subsubsection{RQ3: Design Principle of Proof Language for NTP}
% 
% As elaborated in \cref{sec:minilang}, the design of MiniLang is motivated and guided by the \emph{minimalist principle}: eliminate unnecessary syntax elements and language constructs, distilling declarative proofs to their essential form.
% This adoption of the minimalist principle is embodied in the three strategies in our translation process (\cref{sec:translation}) through the contrast between Isar and MiniLang: Elaboration processes eliminate unnecessary components such as pronouns and connectives; Normalization processes consolidate redundant syntactic approaches that achieve the same logical purpose; and Elimination of tactics removes non-essential tactic applications.
% In the ablation study (\cref{tab:PISA}), the contrast between MiniLang - SH* and Isar shows substantial improvements from the joint application of elaboration and normalization; the contrast between MiniLang and MiniLang - SH demonstrates the effectiveness of tactic elimination.
% These experimental results are sufficient to show,
% \begin{finding}
% \textbf{RA3}: The minimalist principle is an effective guide for designing a proof language for declarative NTP in real-world proof engineering.
% \end{finding}

\subsubsection{RQ 3: Training Corpora for a New Proof Language}
The Minilang training corpus is obtained by translating Isabelle's existing proofs.
The translation process is incomplete, losing $\sim$15\% of proofs, with higher loss rates observed for longer declarative proofs (\cref{fig:translation-distr}). Despite this data reduction, Minilang's experimental results still demonstrate clear advantages due to language improvements. This validates the effectiveness of our translation-based approach.
\begin{finding}
\textbf{RA 3}: Rule-based translation is an effective approach to obtain training corpora for a redesigned proof language.
\end{finding}

\section{Related Works}\label{sec:related-works}

To the best of our knowledge, this work represents the first attempt to improve NTP through comprehensive proof language redesign.
We discuss related work across several areas.

\subsection{Analysis of LLMs' challenges in learning proof languages}

PALM~\cite{PALM} conducts a formative study analyzing the direct causes of failures in a GPT-based NTP system for Coq's proof language Gallina.
One of their key findings is that LLMs often produce proofs with correct high-level structure but struggle with low-level details, justifying the declarative NTP paradigm underlying our work. Additionally, they demonstrate that proof language features --- such as explicit structuring mechanisms like bullets --- significantly impact LLM performance, supporting our approach of improving NTP through language redesign.

\subsection{Proof Representations for Neural Theorem Proving}\label{sec:representations}

%Many works~\cite{Kimina,HOLStep,TacticToe,LLMSTEP,IsarStep} use their own representations mixed with proof scripts, proof states, and/or human solvers' informal thoughts.

Many works~\cite{Kimina,HOLStep,TacticToe,LLMSTEP,IsarStep} design specialized proof representations for training language models on theorem-proving tasks.

One line of work uses representations that combine multiple types of information while retaining the original proof language syntax.
For instance, Kimina~\cite{Kimina} integrates informal proofs with formal proof sketches; TacticToe~\cite{TacticToe} combines proof states with tactic application histories; HOLStep~\cite{HOLStep} and IsarStep~\cite{IsarStep} provide datasets and benchmarks where models work with proof states alongside proof step sequences. LLMSTEP~\cite{LLMSTEP} uses representations combining proof states and file context.

Another line of work transforms proof texts into alternative data structures. Passport~\cite{Passport} represents Coq proofs as trees to capture their hierarchical structure; Paliwal et al.\cite{Paliwal2020} use graph representations to model dependencies between proof steps; CoqGym\cite{CoqGym}, MLFMF~\cite{MLFMF}, and HOList~\cite{HOList} adopt S-expression formats.
%In contrast to these syntactic transformations, our work goes beyond surface-level restructuring by normalizing redundant proof idioms and eliminating expert-oriented features to create a semantically simpler proof language.

Finally, Thor~\cite{Thor} pioneered the use of Sledgehammer to simplify proof languages for declarative NTP. It exhaustively replaces tactics with Sledgehammer to obtain proofs closer to declarative outlines, thereby shifting the burden of tactic reasoning from language models to automated theorem provers.
However, this elimination is incomplete in several aspects.
First, Thor does not normalize Isar's redundant proof idioms and retains Isar's reasoning mechanisms that are unnecessary for declarative NTP --- such as connectives, generic elimination, and calculational reasoning. Second, it does not leverage relevant lemma hints to improve Sledgehammer's premise selection, limiting the effectiveness of proof automation. Third, Thor does not properly handle cases where a single tactic addresses multiple subgoals.
Our work extends Thor's vision of a purely declarative language for NTP by addressing these limitations through comprehensive language redesign and enhanced Sledgehammer*.

% Some works transform the syntactical representations of the language, like Passport's tree representation of Coq~\cite{Passport}, graph representation of HOL~\cite{Paliwal2020}, and the S-expressions adopted by CoqGym~\cite{CoqGym}, MLFMF~\cite{MLFMF}, and HOList~\cite{HOList}. However, they do not change the core (semantics, proof model) of their proof languages.
% 
% Regarding redesigning a proof language to improve NTP's performance, the paper is the first to the best of our knowledge.

%Comparing to existing proof languages, 

\subsection{Prior Neural Theorem Provers on Isabelle}

Leading NTP works on Isabelle and PISA include Thor~\cite{Thor}, Baldur~\cite{Baldur}, and Magnushammer~\cite{Magnushammer}.
Thor has been discussed in \cref{sec:representations}. It employs Google's proprietary models and devices, achieving a success rate of 57.0\% on PISA under a computational budget of at most 300 model queries.
Baldur fine-tunes the LLM Minerva~\cite{Baldur} to generate whole Isabelle/HOL proofs. It also incorporates a repair model that leverages Isabelle's error messages to fix broken proofs. Baldur reaches 65.7\% with pass@64 on PISA.
Magnushammer~\cite{Magnushammer} adopts contrastive learning to target premise selection, the same task as Sledgehammer.
Though this approach does not involve proof generation by LLMs, its combination with Thor reaches the previous state-of-the-art, a success rate of 71\% on PISA.

All aforementioned works use closed-source models, preventing direct performance comparisons under identical configurations. Nonetheless, our ablation study replicates Thor's approach within our experimental framework. Under identical experimental conditions, Minilang achieves ${\sim}20$ percentage points improvement over Thor-style Isar (\cref{tab:PISA}).
Furthermore, under a smaller computational budget than Magnushammer (500s timeout, 16 CPU cores, and up to 8 proof attempts per goal), we achieve a pass@8 success rate of 79.2\% on PISA. Our end-to-end results represent new state-of-the-art performance on the PISA benchmark.

\section{Limitations and Future Work}\label{sec:conclusion}

% A benefit of Minilang's proof model is that its proof state integrates all subgoals and contextual information in a structural way.
% This paper focused on whole proof generation.
% Future work could explore Baldur-style proof repair, or stepwise approaches that could leverage this proof state structure, such as reinforcement learning.

The main trade-off due to Minilang's restriction on language elements is that some goals may require more proof steps, although they are still provable because Minilang is no less proof-theoretically complete than Natural Deduction (Theorem~\ref{thm:completeness}).
Though this trade-off might be unacceptable to human users who prefer efficiency, it aligns well with LLMs' computational strengths: generating lengthy proofs by repeatedly applying learned operations.

While this work implements Minilang in Isabelle/HOL, it is possible to port at least the core of Minilang to other proof assistants like Lean.
This portability stems from Minilang's focus on declarative proof structures, where the notions are general across specific logics and software systems.
As an instance, all Minilang commands in \cref{tab:operations} have correspondences in Lean.
Relying on powerful ATPs, we believe most disparities between proof assistants can be harmonized.
Hopefully, Minilang can serve as a bridge, preventing the long-standing fragmentation in the field of proof assistants from spreading into the field of NTP.
%However, challenges are present.
%First, term language and the core logics are not unifiable. Models still need to learn them.
%Second, a lot of translation rules are required to cover corner cases as many as possible. It also bring a big challenge to the power of the inherent automation of PA.
% 
% In the whole proof generation setting, further work could be done to refine the provided LLM context - the current approach simply takes the most recent items from the same file. Better premise selection could look across multiple files to create contexts with more relevant information.
% 
% We leave these mentioned aspects to our future work.
% 
% 
%  \DS{Don't use unfortunately or the most basic machine learning experiments. If it cannot be shown as a feature, simply say that This paper uses supervised finetuning for the experiments, exploring deeper interactions that could leverage this proof state structure is left as future work.}
% 
% 
% TODO: ????

%\newpage
%
%
%
%
%However, deficiencies in Isabelle's concurrent infrastructure for machine learning impeeded researchers' development on Isabelle-based NTP, which may explain why .
%Nonetheless, this deficiency is addressed by a new infrastructure work in this paper.
%
%
%
%
%
%
%
%
%
%%the practical development of cut-edging AI techniques relies on the support of infrastructures.
%
%Unfortunately
%

\section{Data Availability Statement}

All our infrastructures and data are open sourced.

\vspace{.5em}
{\noindent\centering
\begin{tabular}{@{}p{0.35\linewidth}p{0.62\linewidth}@{}}
$\bullet$\ \ Isabelle REPL infrastructure & \url{https://github.com/xqyww123/Isa-REPL} \\
$\bullet$\ \ Minilang interpreter        & \url{https://github.com/xqyww123/Isa-Mini} \\
$\bullet$\ \ Minilang translator         & \url{https://github.com/xqyww123/Isa-Mini-Translator} \\
$\bullet$\ \ Sledgehammer*         & \url{https://github.com/xqyww123/auto_sledgehammer} \\
$\bullet$\ \ Machine Learning Framework         & \url{https://github.com/xqyww123/MLML} \\
$\bullet$\ \ Translated Minilang Corpus    & \url{https://huggingface.co/datasets/ANTPG/Minilang-AFP-v1} \\
$\bullet$\ \ Language Models         & \url{https://huggingface.co/collections/ANTPG/minilang-oopsla26-models} \\
\end{tabular}
\par}

\bibliographystyle{IEEEtran}
% argument is your BibTeX string definitions and bibliography database(s)
\bibliography{bib}
%
% <OR> manually copy in the resultant .bbl file
% set second argument of \begin to the number of references
% (used to reserve space for the reference number labels box)
%\begin{thebibliography}{1}
%
%\bibitem{IEEEhowto:kopka}
%H.~Kopka and P.~W. Daly, \emph{A Guide to \LaTeX}, 3rd~ed.\hskip 1em plus
%  0.5em minus 0.4em\relax Harlow, England: Addison-Wesley, 1999.
%
%\end{thebibliography}

\appendix
\section{List of Translation Passes}\label{appendix:all_passes}

See \cref{tab:all_passes}.

\begin{table}[p]
\centering
    \caption{Translation passes in our translator from Isar to Minilang, presented in their execution order. LoC = the number of lines of Isabelle/ML code.}
    \label{tab:all_passes}
    \begin{tabular}{c r p{11.3cm}}\toprule
\bf No. & \bf LoC & \bf Description \\\midrule
1. & 83  & Desugaring macro variables like \texttt{?thesis} and \texttt{?case} \\
2. & 76  & Assigning names to anonymous lemmas and intermediate statements. \\
3. & 30  & Converting the commands \isar{also}, \isar{finally}, \isar{moreover}, and \isar{ultimately} into \isar{using}. \\
4. & 42  & A single statement declaration command in Isar can involve attribute modifiers and let-binding declarations. This pass moves the attributes and let-bindings embedded in each statement declaration into separate commands.\\
5. & 64  & Resolving pronouns (macro fact references) like \texttt{this} and \texttt{that}.\\
6. & 39  & Normalizing the commands \isar{with}, \isar{from}, \isar{hence}, \isar{then}, \isar{thus}, and consecutive multiple \isar{using} into a single \isar{using} command.\\
7. & 280 & Fact references in Isar can be by name or by expression pattern. This pass resolves all expression patterns and replaces them with the names of the referenced facts. \\
8. & 44  & Normalizing \isar{subgoal} command into the standard \isar{proof-qed} block. \\
9. & 26  & Checking if all \isar{consider} is used exclusively for case splitting. If not, the translation is unsupported, and an error is raised.\\
10.& 32  & Tactics can appear in many Isar commands (e.g., \isar{by}, \isar{apply_end}, \isar{proof}, \isar{apply}). This pass moves all the tactics to separate \isar{apply} command. \\
11.& 55  & Removing place holder tactics (\texttt{-}). \\
12.& 14  & Normalizing all \isar{using} \textit{facts} commands into tactic ``\texttt{use} \textit{facts} \texttt{in} $\cdots$''. This pass eliminates all \isar{using} commands.\\
13.& 238 & Normalizing \isar{proof-qed} blocks such that each block targets exactly one proof goal in the order the goals appear. \\
14.& 21  & Moving \isar{assume} and \isar{fix} commands to the beginning of each \isar{proof-qed} block. \\
15.& 22  & Normalizing tactics \texttt{induct\_tac} and \texttt{case\_tac} into \texttt{induct} and \texttt{cases}. \\
16.& 55  & Normalizing the tactic \texttt{goal\_cases} into the standard \texttt{proof(cases)-next-qed}. \\
17.& 30  & Collecting information about the fixed hypotheses and variables from the assume and fix commands within each \isar{proof-qed} block. \\
18.& 29  & Normalizing the bracket syntax \isar{\{...\}}\!\!\!\!into the standard \isar{proof-qed} block.\\
19.& 279 & Eliminating tactic combinators and normalizing the tactic applications such that each application affects the leading subgoal only. \\
20.& 42  & Flattening \isar{proof-qed} blocks that contain exactly one nested \isar{proof-qed} block. \\
21.& 393 & Translating the normalized Isar proof into Minilang proof. \\
22.& 31  & Normalizing \minilang{APPLY} (\texttt{use} \textit{facts} \texttt{in} \textit{tactic}) into \minilang{APPLY} (\textit{tactic}) \minilang{WITH} \textit{facts}; Normalizing \minilang{APPLY} (\texttt{insert} \textit{facts}) \minilang{APPLY} (\textit{tactic}) into \minilang{APPLY} (\textit{tactic}) \minilang{WITH} \textit{facts}; Handling other cases relating to \texttt{insert}. \\
23.& 121 & Normalizing tactics \texttt{rule\_tac}, \texttt{erule\_tac}, \texttt{drule\_tac}, \texttt{frule\_tac}, \texttt{rule}, \texttt{drule}, \texttt{frule}, \texttt{erule}, \texttt{intro}, \texttt{elim}, \texttt{dest}, \texttt{standard}, \texttt{unfold\_locales} into \minilang{RULE}. \\
24.& 46  & Normalizing \minilang{RULE} (\texttt{exI}), \minilang{RULE} (\texttt{ex1I}), \minilang{RULE} (\texttt{bexI}) into command \minilang{CHOOSE}. \\
25.& 36  & Applying SH* to eliminate tactics. (The code for SH* is not included.) \\
26.& 25  & Normalizing the remaining tactics \texttt{simp}, \texttt{auto}, \texttt{simp\_all}, \texttt{clarsimp} into the command \minilang{SIMPLIFY}. \\\midrule
\bf \!\!\!\!Total\!\!\! & \!2153 & The transformation from Isar AST to Minilang AST. The parser, printer, evaluator, basic infrastructure, and other facilities are not included.
\\\bottomrule
    \end{tabular}
\end{table}

% that's all folks
\end{document}